\documentclass[english,apj]{emulateapj}
\usepackage[latin9]{inputenc}
\usepackage{graphicx}
\usepackage{amssymb}
\usepackage{esint}

\makeatletter
\providecommand{\tabularnewline}{\\}

\shorttitle{Evolution of binaries near massive black holes}\shortauthors{Antonini~\&~Perets}

\makeatother

\usepackage{babel}

\begin{document}

\title{Secular evolution of compact binaries near massive black holes: \\
Gravitational wave sources and other exotica}

\author{Fabio Antonini$^{1}$ and Hagai B. Perets$^{2,3}$ }

\affil{$^{1}$Canadian Institute for Theoretical Astrophysics, University
of Toronto, 60 George St., Toronto, Ontario M5S 3H8, Canada}

\affil{$^{2}$Harvard-Smithsonian Center for Astrophysics, 60 Garden St.,
Cambridge, MA, USA 02138\\
$^{3}$Technion - Israel Institute of Technology, Haifa, Israel\\
}

\global\long\def\rsun{{\rm \, R_{\odot}}}
\global\long\def\msun{{\rm \, M_{\odot}}}
\global\long\def\kms{{\rm \, km\, s^{-1}}}
\global\long\def\au{{\rm \, au}}
\global\long\def\mpc{{\rm \, mpc}}

\begin{abstract}
The environment near super massive black holes (SMBHs) in galactic
nuclei contains a large number of stars and compact objects. A fraction
of these are likely to be members of binaries. Here we discuss the
binary population of stellar black holes and neutron stars near SMBHs
and focus on the secular evolution of such binaries, due to the perturbation
by the SMBH. Binaries with highly inclined orbits in respect to their
orbit around the SMBH are strongly affected by secular Kozai processes,
which periodically change their eccentricities and inclinations (Kozai-cycles).
During periapsis approach, at the highest eccentricities during the
Kozai-cycles, gravitational wave emission becomes highly efficient.
Some binaries in this environment can inspiral and coalesce at
timescales much shorter than a Hubble time and much shorter than similar
binaries which do not reside near a SMBH. The close
environment of SMBHs could therefore serve as catalyst for the inspiral and
coalescence of binaries, and strongly affect their orbital properties. 
Such compact binaries would be detectable
as gravitational wave~(GW) sources by the next generation of GW detectors
(e.g. advanced- LIGO). About $0.5~$\% of such nuclear
 merging binaries will enter the LIGO observational window while on orbit that are still very eccentric~($e \gtrsim 0.5$).
The efficient gravitational wave analysis  for such systems would therefore require 
the use of eccentric templates.  We also find that binaries 
very close to the SMBH could evolve through a complex dynamical (non-secular) evolution leading to emission of 
several GW pulses during only a few yrs (though these are likely to be rare). Finally, we note that 
the formation of close stellar binaries, X-ray binaries and their
merger products could be induced by similar secular processes, combined
with tidal friction rather than GW emission as in the case of compact
object binaries. 
\end{abstract}


\section{Introduction}

The majority of stars reside in binary and higher multiplicity systems
\citep{rag+10,eva11}, and are observed to exist both in the field
as well as in dense stellar environments such as globular clusters
and galactic nuclei. In dense clusters the dynamical evolution of
binaries can be strongly affected by encounters with others stars.
In the field, encounters with other stars are rare, and do not affect
the binary evolution. Such binaries may still dynamically evolve if
they reside in triple (or higher multiplicity) systems. The inner
binary in a triple can be affected by the perturbation of the third,
outer companion, and evolve through long term secular processes. In
particular, triples in which the relative inclination between the
orbit of the inner binary and the outer binary is high can evolve
through Kozai-Lidov oscillations \citep{1962K,1962L} in which the
eccentricity and inclination of the inner binary significantly change
in a periodic/quasi-periodic manner. 

In this study we explore the evolution of binaries in the environments
of SMBHs which are thought to reside at the center
of most galaxies \citep[e.g. ][]{geb+00,geb+03}; throughout this
paper we focus on galactic nuclei similar to the Galactic center (GC)
of our own galaxy. Close to the center, where the gravitational potential
is dominated by the mass of the SMBH, binaries are bound to the SMBH
and effectively form a triple system, in which the binary orbit around
the SMBH is the outer orbit of the triple. Binaries near a SMBH may
therefore evolve both through gravitational scattering by other stars
in the dense nuclear cluster, as well as through secular evolution
in the SMBH-stellar binary triple system~\citep{ant+10}. Various types of binaries
exist in such environments, including both main-sequence (MS) binaries,
compact object (CO) such as white dwarf (WDs), neutron star (NS) or
stellar black hole (BH) binaries or mixed CO-MS binaries. In this
paper we  focus on CO-binaries, their evolution
into short periods and the way in which they may become potentially
observable GW sources for the advanced laser
interferometer GW observatory (LIGO). Scattering with other stars is also accounted for, but only in the
sense of the lifetime of binaries in the GC, i.e. we do not follow
the evolution of binaries which are disrupted (evaporated) through
encounters with other stars at timescales shorter than the relevant
secular timescales (Kozai evolution). 

In the following we begin with a discussion of binaries in the GC
and their expected properties (Section \ref{sec:Binaries}), we then
discuss the relevant timescales for the evolution of stars and binaries
in galactic nuclei (Section \ref{sec:time-scales}). In Section \ref{sec:GW-sources}
we discuss the secular evolution of binaries near a SMBH, and their
potential coalescence due to a combined Kozai cycles-GW loss mechanism;
we then calculate the type and properties of the GW sources which
form through this process. The secular evolution of other types of
binaries is briefly discussed in Section \ref{sec:KCTF} . We summarize
our results in Section~\ref{sec:summary}.

\section{Binaries in the Galactic center}

\label{sec:Binaries}

At present, little is known about binaries in the GC. Massive binaries
can be currently detected only by observations of stellar variability
\citep{ott+99,dep+04,mar+06b,raf+07}, which can mainly detect eclipsing
or ellipsoidal variable binaries \citep{raf+07}. The fraction of
such binaries among O-stars outside the GC is $2-11\,\%$ \citep{gar+80,raf+07}.
\citet{raf+07} conducted a variability survey of the central $5"\times5"$
of the GC and found one such binary  \citep[also observed by][]{ott+99,dep+04,mar+06b,raf+07}
among 15 massive stars ($7\%$); a fraction consistent with the Galactic
one. Other binary candidates were also found: a colliding wind binary
\citep{raf+07} and another eclipsing binary \citep{pee+07}. Although
these data are still insufficient for drawing strong conclusions about
the total massive binary fraction in the GC, they do suggest that
it is comparable to that observed outside the GC, among massive binaries.
For reference, the massive binary fraction in the Solar neighborhood
is very high ($>70\,\%$; \citealt{gar+80,mas+98,kou+07,eva11,kob+07}),
and most massive binaries have semi-major axes of up to a few AU.
Given our poor understanding of binary formation, it is unknown whether
star formation in the unusual GC environment results in similar binary
properties. 

No observational data exist regarding low mass binaries in the GC.
Little is known about compact binaries in the GC, though the observed
overabundance of X-ray sources in the central pc \citep{mun+05} suggests
they are not rare but their properties and origin are not known. 

We focus on binaries very close to the SMBH, where secular Kozai evolution
affects the binary dynamics. However, this population is continuously
destroyed through various processes~(evaporation, mergers, disruption).
We therefore first discuss the binary population outside the central
region near the SMBH ($>\sim2\,$pc), which serves as a continuous source
term repopulating the binaries in the central region. We then discuss
the processes which can repopulate binaries in the central region,
both from populations outside and from in-situ formation of new binaries.

\subsection{Binaries outside the radius of influence of the SMBH}

In the following we discuss the binary population outside the central
region near the SMBH ($>\sim2\,$pc). This binary population serves
as a source term for binaries in the central region on which we focus,
which are continuously destroyed through various processes. 

\begin{figure}
\begin{tabular}{c}
\includegraphics[clip,width=1\columnwidth]{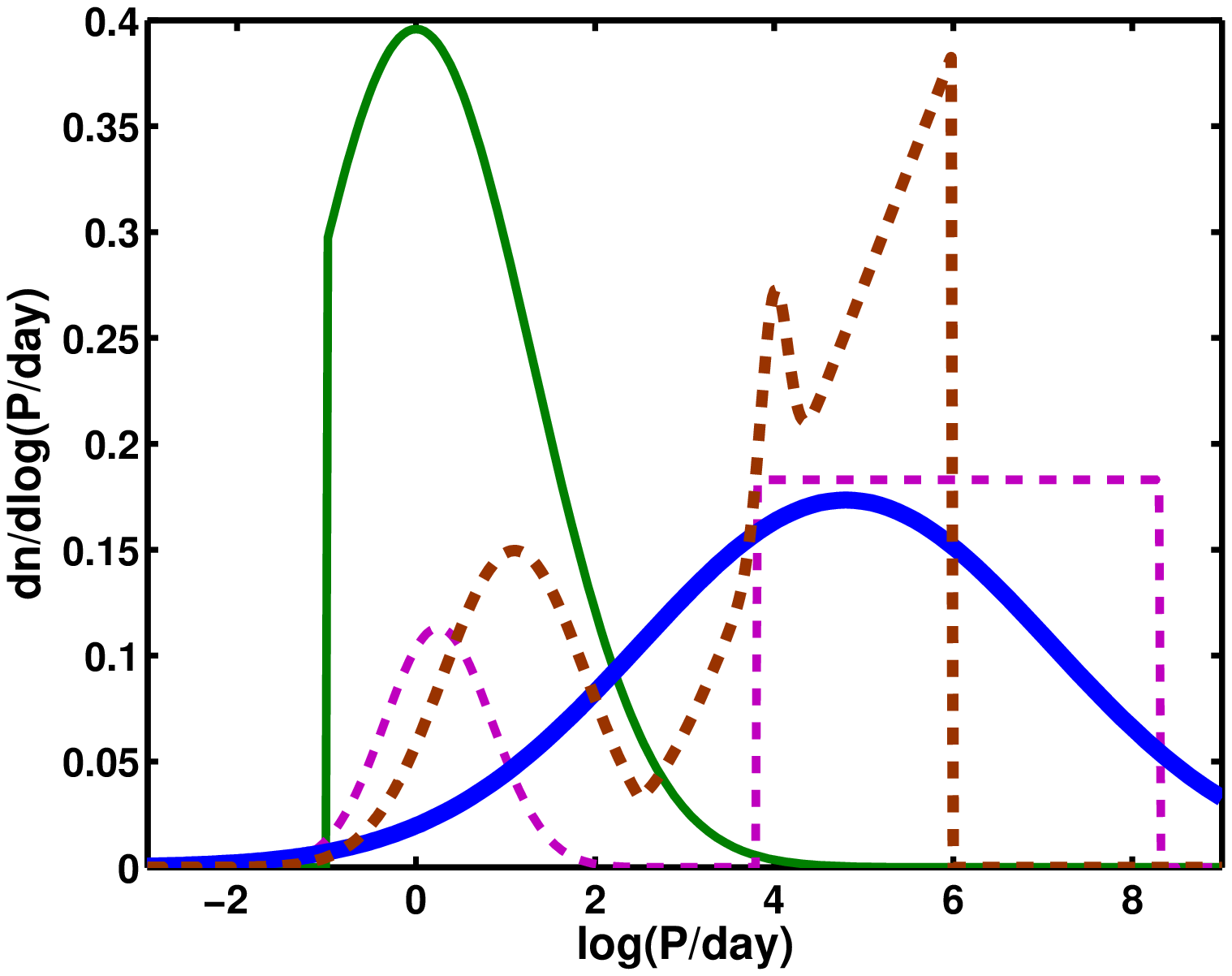}\tabularnewline
\end{tabular}

\caption{\label{f:bin-period} The period distribution of primordial binaries
(without dynamical effects). Main sequence stars (solid thick line)
distribution taken according to observations \citet{duq+91}. WD distribution
(thin dashed line) is taken according to the theoretical model of
\citet{wil+04a}. NS binaries distribution (thin solid line) is taken
according to distribution of binary pulsars in the ATNF database \citep[at http://www.atnf.csiro.au/research/pulsar/psrcat/ ; ][]{man+05}.
BH binaries distribution (thick triple peaked dashed line) is taken
according to the theoretical model of \citet{bel+04a}. }

\end{figure}

The binary fraction and semi-major axis distribution of observed main
sequence stars in the solar neighborhood have been analyzed in many
binary surveys (e.g. \citet{duq+91,lad06} and references within).
For compact objects the observational data is still lacking. However,
many theoretical analysis have been done using stellar evolution and
binary synthesis codes. Most of these studies focused on close binaries,
usually in the context of CVs, x-ray binaries or gravitational waves
sources (which are relatively short lived and compose only a very
small fraction of the binary population), and did not analyze the
general distribution, which is needed for our analysis. Nevertheless,
the general semi-major axis distribution of COs binaries have been
found for isolated (i.e. not taking into account dynamical effects
which could be important in dense stellar environments) WD-MS binaries
\citep{wil+04a} and BH binaries \citep{bel+04a}. For NSs, one can
use the observed pulsar binaries population as found in the ATNF pulsar
catalog \citep[at http://www.atnf.csiro.au/research/pulsar/psrcat/ ; ][]{man+05}.
The initial period distribution we consider for the different binary
populations is shown in Fig. \ref{f:bin-period}. These distributions
do not take into account dynamical evolution of binaries in the dense
environment near the SMBH, mainly important in evaporating wide binaries
(so called soft binaries; \citealt{heg75}) due to distant stellar
encounters. They also do not account for the long term evolution of
close compact object binaries due to GW emission. We account for these
evolutionary aspects using simplified methods. 

The primordial binary fraction (not taking into account dynamical
evolution such as evaporation of the binaries due to encounters),
$f_{bin}$ , is different for the different binary types, MS, WD,
NS or BH binaries. It depends both on the progenitors binary fraction,
$f_{bin}^{pro}$ (i.e. BHs and NSs progenitors are high mass stellar
binaries whereas WD binary progenitors have lower masses) and fraction
of surviving binaries, $f_{surv}$, after going through their evolutionary
stages . 

The binary fraction of massive stars ($f_{bin}^{pro}=0.7$) is more
than two times that of lower mass stars ($f_{bin}^{pro}=0.3$; \citealt{kob+07,lad06}).
Stellar evolution further changes these fractions. Thus, for obtaining
$f_{bin}$ we either take the observed binary fractions in the Solar
neighborhood (in the cases of MS binaries and NS binaries) or use
the theoretical models for the survivability of the different binaries
\citep{wil+02,wil+04a,bel+04a} to find $f_{surv}$, and multiply
it by the progenitors binary fractions $f_{bin}^{pro}$ (i.e. $f_{bin}=f_{bin}^{pro}f_{surv}$).
We finally get $f_{bin}=0.7,\,0.3,\,0.3,\,0.07,\,0.1$ (for high and
low mass MS stars, WDs, NSs and BHs respectively) as the evolution
of lower mass stars (and WDs) changes the semi-major axis distribution
of the binaries but does not result in their disruption, whereas the
evolution of BHs and NSs results in most cases in the binary disruption.
We adopt these initial binary fractions throughout our calculations. The fractions of CO binaries used here already account for the initial high binary fraction and the stellar evolution of these binaries. However, these binary fractions are also expected to be reduced due to dynamical encounters
with other stars  and through
coalescence due to GW emission (for compact objects) and tidal friction. In the following calculations (specifically in section \ref{sec:merger_frac}) we assume the initial CO binary fractions ($f_{bin}$) in the GC are similar to those obtained here, where the effects due to dynamical encounters and Kozai evolution of compact binaries are accounted for in a simplified manner, as we discussed later on. Evolution due to tidal frction which is important for MS stars will be discussed elsewhere, and is only shortly discussed here.

\begin{figure*}
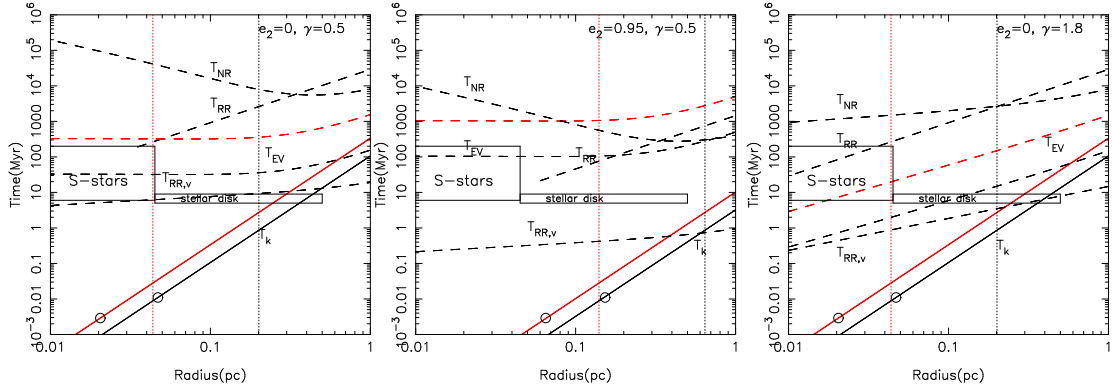

\centering{}\includegraphics[width=2.in,angle=270.]{Figure2a.eps}\includegraphics[width=2.in,angle=270]{Figure2b.eps}
\includegraphics[width=2.in,angle=270]{Figure2c.eps} \caption{\label{fig:times}Time scales at the Galactic center (dashed lines)
compared to the Kozai period ($T_{{\rm K}}$; continue lines) for
a $(10+10)~M_{\odot}$~(red lines) and $(1+1)~M_{\odot}$~(black
lines) binaries with $a_{1}=1~$AU. Rectangular regions show the age
and radial extend of the S-stars and disk of WR/O stars. Dotted lines
give the radii at which Kozai cycles are fully suppressed by relativistic
precession in the inner binary. Open circles indicate the radius where
the time-scale of periapsis precession due to tidal bulges equals
$T_{{\rm K}}$ assuming an inner eccentricity $e_{1}$ such that the
closest approach between the stars is twice the sum of their radii,
i.e., $e_{1}=1-2\times\left(R_{0}+R_{1}\right)/a_{1}$.}

\end{figure*}

\subsection{The binary fraction near a SMBH}

As discussed above, the binary fraction in the GC is unknown; and
the current observational data is insufficient to prove any meaningful
constraints. Very few theoretical studies explored the role and evolution
of binaries near SMBH. \citet{bah+77} briefly discussed the effect
of binaries on the dynamics of stars in a cusp around a SMBH. They
showed that their effect on the cusp dynamics is small, since the
vast majority of these binaries are soft, and the total energy which
they can transfer to the cusp stars is small in respect to their orbital
energy (in the potential of the SMBH). \citet{per+08c} suggested that
binaries in the observed stellar disk in the GC can have a non-negligible
effect on the evolution of the cold disk (see also \citealt{cua+08}).
Binaries can also be important for a fast deposition of stars close
to a SMBH, through binary disruption (\citealt{hil88,gou+03,per+07,mad+09}).
These studies did not discuss the binary population very close to
the SMBH; the only study to explore the general population of binaries
near SMBH was done by \citet{hop09}, even though this pioneering study
accounted only for regular relaxation processes. A full study of the
binary population near SMBH is highly desirable. However, this is beyond
the scope of this study which mainly focuses on the secular evolution
of binaries due to the perturbation by a SMBH. Here we assume a simple
model in which the binary population in galactic nuclei follows their
population in the field, but we also account for their evaporation
through taking their effective lifetimes to be the evaporation timescale
at the distance from the SMBH where they reside. We discuss several
possible replenishment of binaries in Section \ref{sub:GW-rates}.

\section{Time scales}

\label{sec:time-scales}

In the dense stellar environment of a galactic nucleus, otherwise
rare dynamical processes can take place and affect the evolution of
the stellar population near the center.  Here we compute the time-scale
associated with such processes in the Galactic center and compare
it to the Kozai time-scale associated with the secular evolution of
binaries due to the perturbations by a central SMBH.

We consider stellar-mass binaries with components of masses $m_{0}$
and $m_{1}$ orbiting a super-massive black hole~(SMBH) of mass $M_{\bullet}$.
We denote the eccentricities of the inner and outer binary, respectively,
as $e_{1}$ and $e_{2}$, and semi-major axes $a_{1}$ and $a_{2}$.
We also define $g_{1}$ as the argument of periapsis of the inner
binary relative to the line of the descending node, and $I$ as the
mutual orbital inclination of the inner binary with respect to the outer
orbit.

The mass density near the SMBH was modeled using the broken power-law
density profile: \begin{equation}
\rho(r)=\rho_{0}\left(\frac{r}{r_{0}}\right)^{-\gamma}\left[1+\left(\frac{r}{r_{0}}\right)^{2}\right]^{(\gamma-1.8)/{2}}\label{den}\end{equation}
with $\gamma$ the inner slope index and $r_{0}=0.5$pc. Setting
$\rho_{0}=5.2\times10^{5}{\rm M_{\odot}pc^{-3}}$ this gives a good
fit to the observed space density at the Galactic center outside the
core, normalized at 1 pc \citep{S:09}. For the slope of the inner
density profile we assumed two different values: $\gamma=0.5$, representative
of the observed distribution of stars at the Galactic center \citep{buc+09,do+09,bar+10},
and $\gamma=1.8$ corresponding approximately to a nearly-relaxed
configuration of stars near a dominating point-mass potential \citep{ale05}.
We set $M_{\bullet}=4\times10^{6}M_{\odot}$, similar to the mass
of the SMBH at the Galactic center \citep{ghe+08,gil+09}.

\emph{Kozai timescale:} The Kozai time-scale can be written in terms
of the masses of the three bodies, the eccentricity of the outer binary
and their semi-major axes as: \begin{eqnarray}
T_{{\rm K}} & \approx & \frac{2~P_{{\rm 2}}^{2}}{3\pi~P_{{\rm {1}}}}\left(1-e_{2}^{2}\right)^{3/2}=1.1\times10^{5}{\rm yr}\left(\frac{a_{2}}{0.1{\rm pc}}\right)^{3}\times\label{kp}\\
&  & \left(\frac{M_{\bullet}}{4\times10^{6}M_{\odot}}\right)^{-1}\left(\frac{M_{b}}{2M_{\odot}}\right)^{1/2}\left(\frac{a_{1}}{1{\rm AU}}\right)^{-3/2}\left(1-e_{2}^{2}\right)^{3/2}~,\nonumber \end{eqnarray}
where $P_{{\rm 1}}$ is the period of the inner binary, $P_{{\rm 2}}$
the period of the outer orbit, and $M_b=m_0+m_1$.

\emph{Extra sources of periapsis precession:} Kozai Cycles can be
suppressed by additional sources of apsidal precession. The most relevant processes to consider are relativistic precession
and, in the case of stellar binaries, precession due to the tidal  bulge raised 
on each of the inner binary's components. Writing the timescale
of relativistic precession as: \begin{eqnarray}
T_{{\rm GR}} & = & \frac{a_{1}^{5/2}~c^{2}~(1-e_{{\rm 1}}^{2})}{3G^{3/2}M_{{b}}^{3/2}}\nonumber \\
&  & =1.9\times10^{6}{\rm yr}\left(\frac{M_{{b}}}{2M_{\odot}}\right)^{-3/2}\left(\frac{a_{1}}{1{\rm AU}}\right)^{5/2}\left(1-e_{1}^{2}\right),\label{eq:tgr}\end{eqnarray}
with $G$ the gravitational constant and $c$ the speed of light,
it can be shown that general relativistic precession in the inner
binary stops the Kozai oscillations when \citep{hol+97,bla+02}: \begin{eqnarray}
\frac{a_{2}}{a_{1}} & < & 4200\left(\frac{a_{1}}{1{\rm AU}}\right)^{1/3}\left(\frac{M_{b}}{2M_{\odot}}\right)^{-2/3}\nonumber \\
&  & \times\left(\frac{M_{\bullet}}{4\times10^{6}{\rm M_{\odot}}}\right)^{1/3}\left(\frac{1-e_{1}^{2}}{1-e_{2}^{2}}\right)^{1/2}.\end{eqnarray}

Tidal bulges always tend to promote periapsis precession and therefore
suppress Kozai cycles. The time scale of periapsis precession due
to non dissipative tides is \citep{kis+98,2001EK}: \begin{eqnarray}
\centering{T}_{{\rm Tide}} & = & \frac{8a_{1}^{13/2}}{15(GM_{b})^{1/2}}\frac{(1-e_{1}^{2})^{5}}{8+12e_{1}^{2}+e_{1}^{4}}\nonumber \\
&  & \times\left[\frac{m_{1}}{m_{0}}k_{0}R_{0}^{5}+\frac{m_{0}}{m_{1}}k_{1}R_{1}^{5}\right]^{-1},\end{eqnarray}
with $R_0$ and $R_1$ the stellar radii and $k_{0}$ and $k_1$ the tidal Love numbers.
Assuming equal objects in the binary and $k_{0}=k_{1}=0.01$, we
find: \begin{eqnarray}
{T}_{{\rm Tide}} & = & 1.3\times10^{12}{\rm yr}\left(\frac{a_{1}}{1{\rm AU}}\right)^{13/2}\left(\frac{M_{b}}{2M_{\odot}}\right)^{-1/2}\left(\frac{R_{0}}{1R_{\odot}}\right)^{-5}\label{eq:ttide}\\
&  & \times\frac{(1-e_{1}^{2})^{5}}{8+12e_{1}^{2}+e_{1}^{4}}~,\end{eqnarray}
where the stellar radii are obtained using the expression \citep{han+04}:
\begin{eqnarray}
R=R_{\odot}\left(\frac{M}{M_{\odot}}\right)^{0.75}~.\end{eqnarray}
Comparing Equations~(\ref{eq:tgr}) and (\ref{eq:ttide}) it can
be seen that in general GR precession is initially the most rapid
mechanism inducing apsidal motion, but, for very large eccentricities,
when, for instance, the system approaches the maximum of a Kozai cycle,
one expects tides to become important to the evolution. Consequently,
for stellar binaries with relatively large semi-major axes, very large
eccentricities (eventually resulting in mass transfer or even collisions)
could be avoided due to apsidal precession induced by tidal bulges.

\emph{Binary evaporation:} In a dense environment, binaries may evaporate
due to dynamical interactions with field stars if \begin{equation}
|E|/({M_{{b}}\sigma^{2}})\lesssim1,\end{equation}
with $E$ the internal orbital energy of the binary, and $\sigma$
the one-dimensional velocity dispersion of the stellar background.
The evaporation timescale is given by
\citep{bin+87}: \begin{eqnarray}
T_{{\rm EV}}=3.2\times10^{7}~{\rm yr}\left(\frac{M_{{b}}}
{2~M_{\odot}}\right)\left(\frac{{\sigma}}{306~{\rm kms^{-1}}}\right)\left(\frac{m}{M_{\odot}}\right)^{-1}\nonumber \\
\times\left(\frac{a_{1}}{{\rm 1~AU}}\right)^{-1}\left(\frac{{\rm ln}\Lambda}
{15}\right)^{-1}\left(\frac{{\rho}}{2.1\times10^{6}M_{\odot}{\rm pc^{-3}}}\right)^{-1}~~~~\label{eq:TEV}\end{eqnarray}
where ${\rm ln}\Lambda$ is the Coulomb logarithm, $\sigma$ and
$\rho$ are the 1D velocity dispersion and mass-density, and $m$
is the mass of a typical object in the system assumed to be solar
mass stars in what follows. The velocity dispersion, $\sigma$, is
calculated from the Jeans equation: \begin{equation}
\rho(r)\sigma(r)^{2}=G\int_{r}^{\infty}dr'r'^{-2}\left[M_{\bullet}+M_{\star}(<r')\right]\rho(r'),\label{jeans}\end{equation}
with $M_{\star}(<r)$ the total mass in stars within $r$.

\emph{Two-body relaxation:} Assuming equal-mass stars and an isotropic
velocity distribution, the local two-body (non resonant) relaxation
time is defined as \citep{spi87} : \begin{eqnarray}
T_{{\rm NR}} & = & 1.6\times10^{10}~{\rm yr}\left(\frac{{\sigma}}{306~{\rm km\ s^{-1}}}\right)^{3}\left(\frac{m}{M_{\odot}}\right)^{-1}\label{eq:nr}\\
&  & \left(\frac{{\rm ln}\Lambda}{15}\right)^{-1}\left(\frac{{\rho}}{2.1\times10^{6}M_{\odot}{\rm pc^{-3}}}\right)^{-1}.\nonumber \end{eqnarray}
Setting $\gamma=0.5$ in the density model of Equation (\ref{den}),
this time results somewhat  larger than the age of the Galaxy ($\sim10^{10}~$yr)
when computed at the radius of influence ($\sim3~$pc) of the SMBH
\citep{mer10}.

\emph{Resonant relaxation:} In the dense stellar environment near
a SMBH, as long as the relativistic precession time scale is much
longer than the orbital period, the mechanism that dominates the scattering
of stars onto high-eccentricity orbits is resonant relaxation. Because
in the potential of a point-mass the orbits are fixed ellipses, perturbations
on a test particle are not random but correlated. The residual torque
$|{\bf Q}|\approx\sqrt{N}Gm/r$, exerted by the $N$ randomly oriented,
orbit-averaged mass distributions of the surrounding stars, induces
coherent changes in angular momentum $\Delta{\mathbf L}={\mathbf Q}t$ on
timescales $t\lesssim T_{coh}$. Where the coherent time $t_{coh}$ is
fixed by the mechanism that most rapidly causes the orbits to precess
(e.g, mass precession, relativistic precession). The angular momentum
relaxation time associated with resonant relaxation is: \[
T_{{\rm RR}}\approx\left(\frac{L_{c}}{\Delta \mathbf{L}_{coh}}\right)^{2}T_{coh}~,\]
where $L_{c}\equiv\sqrt{GM_{\bullet}a}$ is the angular momentum
of an orbit with radius $r\approx a$ and $|\Delta L_{coh}|\sim|{\bf Q}T_{coh}|$
is the accumulated change over the coherence time. Assuming that the
precession is determined by the mean field of stars (i.e., mass precession),
the angular momentum relaxation time is \citep{rau+96}: \begin{equation}
T_{{\rm RR}}\approx9.2\times10^{8}{\rm yr}\left(\frac{M_{\bullet}}{4\times10^{6}
{\rm M_{\odot}}}\right)^{1/2}\left(\frac{a_{2}}{{\rm 0.1pc}}\right)^{3/2}\left(\frac{m}{{\rm M_{\odot}}}\right)^{-1}.\label{eq:rr}\end{equation}
At small radii relativistic precession becomes efficient at suppressing
resonant relaxation and the gravitational perturbations are dominated
by classical (two-body) non-resonant relaxation. The value of the angular momentum
(the Schwarzschild barrier) at which this transition occurs is \citep{mer+11}:
\begin{eqnarray}
(1-e_{2}^{2})_{{\rm SB}} & \approx & 6\times10^{-3}\left(\frac{C_{{\rm SB}}}{0.7}\right)^{2}\left(\frac{a_{2}}{{\rm 0.1~pc}}\right)^{-2}\nonumber \\
&  & \left(\frac{M_{\bullet}}{4\times10^{6}{M_{\odot}}}\right)^{4}\left(\frac{m}{{M_{\odot}}}\right)^{-2}\left(\frac{N}{10\times10^{4}}\right)^{-1}~,~~~~~\end{eqnarray}
where $N$ is the number of stars within radius $a_{2}$ and $C_{{\rm SB}}$
is a constant of order of unity.

\emph{Vector resonant relaxation:} An additional fundamental process
in changing the external orbit of the binary is vector resonant relaxation~(VRR).
The external orbital plane is randomized non-coherently on the vector
resonant relaxation timescale \citep{hop+06a}: \begin{eqnarray}
T_{{\rm RR,v}}\approx7.6\times10^{6}\left(\frac{M_{\bullet}}{4\times10^{6}{\rm M_{\odot}}}\right)^{1/2}\left(\frac{m}{M_{\odot}}\right)^{-1}\nonumber \\
\times\left(\frac{a_{2}}{{\rm 0.1pc}}\right)^{3/2}\left(\frac{N}{6000}\right)^{-1/2}{\rm yr}~.~~~~~~~~~~~~~~\label{eq:rrv}\end{eqnarray}

\begin{figure*}
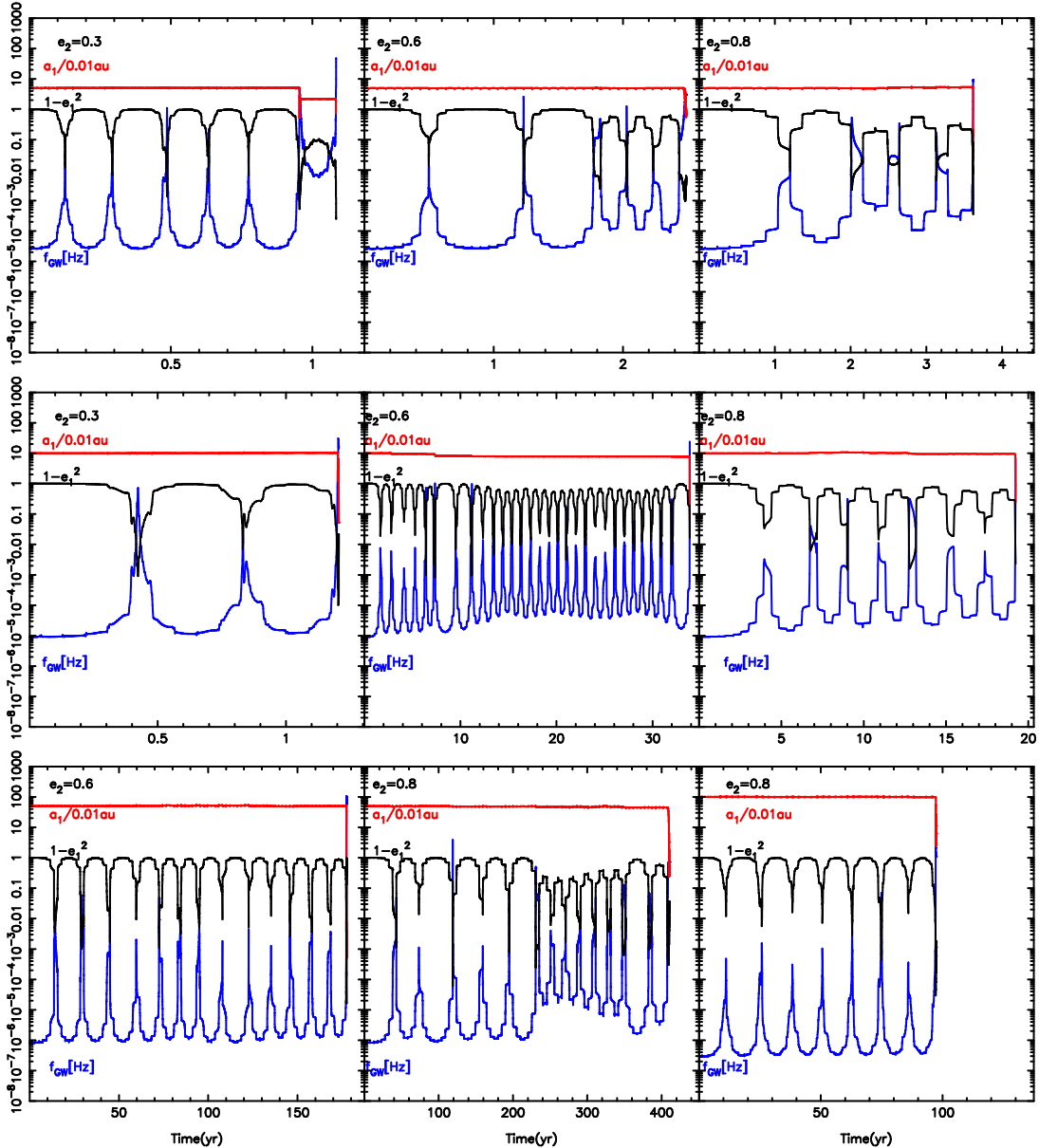

\centering{}\includegraphics[width=5.7in]{Figure3a.eps}\\
\includegraphics[width=5.7in]{Figure3b.eps} \\
\includegraphics[width=5.7in]{Figure3c.eps}
\caption{\label{example1} Time evolution of the gravitational wave frequency~($f_{{\rm GW}}$),
eccentricity~($e_{1}$) and semi-major axis of binary black holes
orbiting a SMBH. In these integrations the distance of closest approach
of the binary to the SMBH is $r_{{\rm per}}=3.3\times r_{{\rm bt}}$,
with $r_{{\rm bt}}$ the tidal disruption radius of the binary. We
set the initial parameters to $e_{1}=0$, $m_{0}=m_{1}=10~M_{\odot}$,
$I=88^{\circ}$, $g_{1}=0^{\circ}$ and adopt different values for
the orbital eccentricity $e_{2}$ and inner semi-major axis $a_{1}$.
The nominal merger timescales of the binaries in isolation are $\sim10^{14}~$yr
($a_{1}=1~$AU; bottom right panel), $\sim10^{13}~$yr ($a_{1}=0.5~$AU;
two bottom  left most panels), $\sim10^{10}~$yr ($a_{1}=0.1~$AU;
middle panels) $\sim10^{9}~$yr ($a_{1}=0.05~$AU; top panels). The
presence of an external perturber, the SMBH in this case, causes periodic
variation of the inner binary eccentricity greatly reducing the merger
time-scales of these systems. }
\end{figure*}

\begin{figure}
\centering{}~
\includegraphics[width=3.2in]{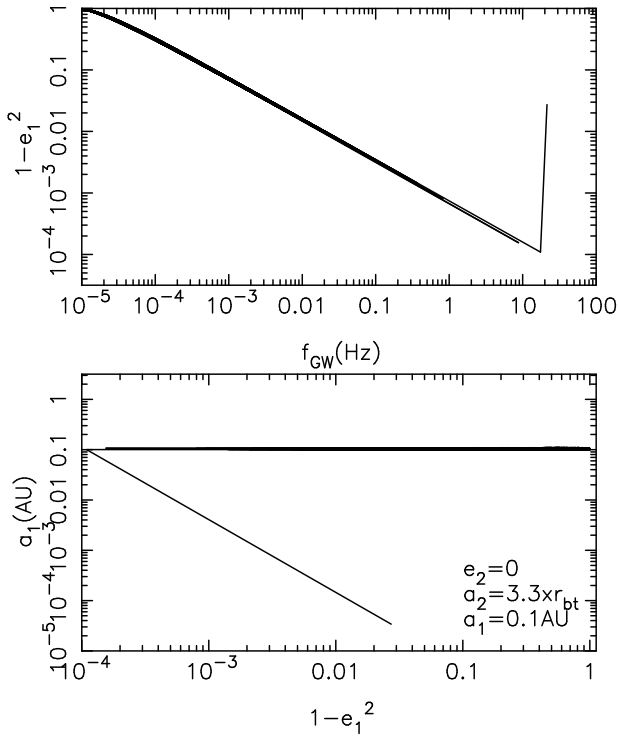}
\caption{\label{example2} Evolution onto the 
$e_1-f_{\rm GW}$ and  $a_1-e_1$  planes of a stellar black hole binary 
orbiting a SMBH on a circular orbit of radius $20~$AU with $I=88^{\circ}$. 
This system enters the
LIGO observational  band with $e\sim1$ and merges in $\sim20$~yr after many periodic variations~($\sim 100$)  of the
inner eccentricity have already occurred.}
\end{figure}

\begin{figure*}
\centering{}\includegraphics[width=2.5in,angle=270.]{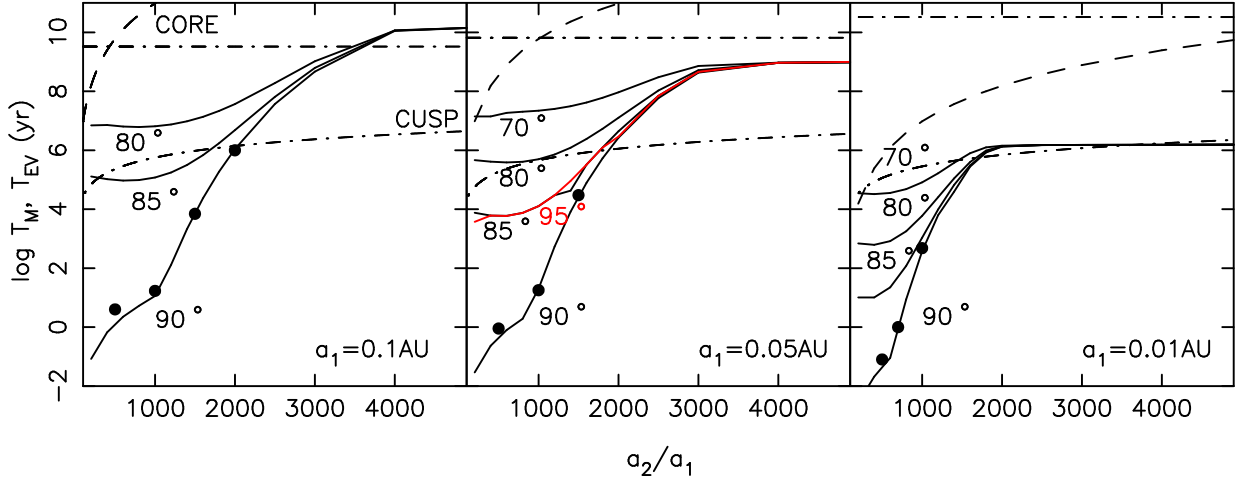} 
\caption{\label{ta1} Merger time~(continue lines) and evaporation time~(dot-dashed
lines) of stellar black hole binaries near a SMBH. The evaporation
time was computed under the assumption of either a cusp or a core
in the density of stars near the center. Highly inclined systems close
to the central black hole merge in a time typically much shorter than
the timescale over which they would evaporate due to close encounters
with field stars. Dashed lines give the inspiral time of the compact
binary into the SMBH due to GW loss. Here we used $m_{0}=m_{1}=10~M_{\odot}$,
$e_{1}=e_{2}=0.1$, and $g_{1}=0^{\circ}$. Black filled circles were obtained
by direct 3-body integrations including PN corrections up to order
2.5.}
\end{figure*}

\begin{figure}
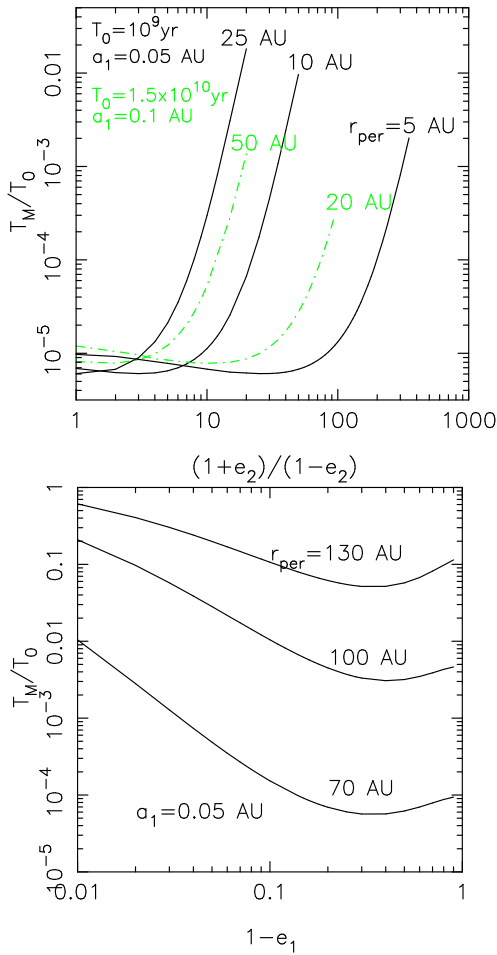

~~~~\centering{}\includegraphics[width=2.5in,angle=270.]{Figure6a}  \includegraphics[width=2.45in,angle=270.]{Figure6b}
\caption{\label{ta} Upper panel: merger time (in units of the merger time of
the binary evolved in isolation) versus $r_{{\rm apo}}/r_{{\rm per}}=(1+e_{2})/(1-e_{2})$.
At large semi-major axis, relativistic precession ''de-tunes'' the
secular Kozai effects and decreases the maximum eccentricity achieved
during a cycle. This results in longer merger times. Bottom panel:
merger time as a function of the internal eccentricity of the compact
binary. For $e_1\gtrsim0.7$, the effect of the Kozai resonance on
the merge time becomes progressively less important as $e_{1}$ increases.
In these computations we used $m_{0}=m_{1}=10~M_{\odot}$, $e_{1}=0.1$,
$I=85^{\circ}$ and $g_{1}=0^{\circ}$.}
\end{figure}

VRR changes the direction but not the magnitude
of the external orbit angular momentum. The randomization process
of the external orbital plane can possibly produce variations in the
mutual inclination of the binary orbit with respect to its orbit around
the SMBH. 

In Figure~2 these different timescales are compared for the case
of a circular orbit and for an eccentric orbit with $e_{2}=0.95$.
In this latter case the corresponding timescales are obtained by multiplying
the quantities of Equations (\ref{eq:nr}),(\ref{eq:rr}) and (\ref{eq:rrv})
by a factor $(1-e_{2})$.  To compensate for the fact that when the external eccentricity is large
the binary spends most of the time at apoapsis, we divided the evaporation time
given in Equation~(\ref{eq:TEV}) by a factor $\sqrt{1-e_2^2}$.
At the radii of interest, for either the cusp and the core model, the Kozai timescale results much shorter
than any other timescale associated with perturbations from dynamical
interaction with field stars. VRR is
the most rapid mechanism associated with gravitational interaction
with field stars. For $a_{2}\lesssim0.1~$pc, since $T_{{\rm RR,v}}\gg T_{K}$,
VRR might lead to an increase of the number of binaries undergoing
Kozai resonances with the SMBH: large inclined binaries will have
already experienced multiple Kozai cycles and eventually merge before
VRR could change appreciably their orbits, while low inclined binaries
could move to high inclinations via VRR and enter the Kozai regime.
The degree to which the binary dynamics depend on VRR and other dynamical
processes is a complicated topic that would most likely require direct
$N$-body simulations describing dense clusters of stars around SMBHs.
In what follows, as a first approximation, we neglect VRR and use
the time for binary evaporation ($T_{{\rm EV}}$) as the timescale
of reference to define the parameter space where our integrations
are valid. 

From Figure~2 we see that within $\sim0.5~$pc, i.e., within the
limiting radius of the stellar disk at the Galactic center, most of
the binaries with initially high inclined orbits would have performed
multiple Kozai cycles during their life time. At $\sim0.2~$pc, Kozai
resonances are completely suppressed by relativistic precession in
the inner binary. However, since increasing the external orbital eccentricity
reduces the Kozai period but leaves the timescale for relativistic
precession unchanged, for a given orbital radius it is always possible
to choose $e_{2}$ in order to have $T_{{\rm K}}>>T_{{\rm GR}}$.
For instance, setting $e_{2}\gtrsim0.9$, relativistic precession
becomes important only at radii of order parsec (middle panel in the
figure). However, very large eccentricities could result in a very
short distance of closest approach of the binary to the SMBH, possibly
resulting in the tidal break up of the system. This occurs when the
binary approaches the SMBH within at a distance \citep{mil+05}: \begin{eqnarray}
r_{{\rm bt}} & \sim & \left(\frac{M_{\bullet}}{M_{b}}\right)^{1/3}a_{1}\label{rbt}\\
&  & \approx120~a_{1}\times\left[\frac{M_{\bullet}}{4\times10^{6}M_{\odot}}\frac{2M_{\odot}}{M_{b}}\right]^{(1/3)}.\nonumber \end{eqnarray}

In this section, we have assumed that the perturbers are solar mass
objects, but, we note here that gravitational scattering can be dominated
by stellar black holes. This can be the case if the age of the nucleus
is larger than the mass-segregation time scale of the heavy component.
Assuming standard initial mass functions, in a relaxed model the mass
in black holes within $0.01~$pc from SgrA{*} is $N_{{\rm BH}}\approx10^{2}$
\citep{fre+06,hop+06a} and Gravitational perturbations will be dominated
by stellar black holes. In this case the time scale for evaporation
will be somewhat  shorten than that implied in Figure~2
within $1~$mpc, but it will be essentially unaffected at larger radii.
In the case of a shallow density cusp around a massive black hole
(i.e., $\gamma\lesssim0.5$) dynamical friction falls essentially
to zero inside the core due to the absence of stars moving more slowly
than the local circular velocity~\citep{ant+12}. This results in
much lower densities of stellar remnants at small radii with respect
to collisionally relaxed models~\citep{gm+12}. The distributed mass in this case
will be therefore dominated by stars at all radii, and gravitational
scattering from black holes can be neglected.

\section{Gravitational wave sources}

\label{sec:GW-sources}

In the following we address the possibility that the merger time of
compact binaries residing in galactic nuclei can be significantly
accelerated due to their interaction with a central SMBH. Some systems
that otherwise would have no chance of merging before being disrupted
by gravitational encounters with background stars~(or stellar remnants)
will actually merge due to combined effect of Kozai cycles and gravitational
wave loss. For an isolated binary, the gravitational wave merger timescale
can be approximated as \citep{pet64}: \begin{eqnarray}
T_{0} & \simeq & \frac{3}{85}\frac{a_{1}}{c}\left(\frac{a_{1}^{3}c^{6}}{G^{3}m_{0}m_{1}M_{b}}\right)(1-e_{1}^{2})^{7/2}\label{tmer}\\
& = & 1.2\times10^{10}{\rm yr}\left(\frac{2\times10^{3}
M_{\odot}^{3}}{m_{0}m_{1}M_{b}}\right)\frac{a_{1}}{0.1~{\rm AU}}(1-e_{1}^{2})^{7/2}~.\nonumber \end{eqnarray}
Given the strong dependence of $T_{0}$ on the eccentricity, the
induced oscillations in the binary orbital eccentricity via secular
Kozai processes can clearly decrease dramatically the merger timescale
\citep[e.g. ][]{tho+11}. The GW signal emitted by such compact binaries
will be completely dominated by repeated pulses emitted during periapsis
passages and at high eccentricities where the GW frequency is maximized
\citep{gou11}.

\subsection{$N$-body integrations}

In this section we describe high accuracy $N$-body simulations that
were used to study the evolution of a limited selection of systems
representing binaries of BHs orbiting a SMBH. These simulations were
carried out using the  AR-CHAIN integrator~\citep{mik+08}, which includes
PN corrections to all pairwise forces up to order PN2.5. The code
employs an algorithmically regularized chain structure and the time-transformed
leapfrog scheme to accurately trace the motion of tight binaries with
arbitrarily large mass ratios. We considered circular and equal mass
binaries with components of masses $10~M_{\odot}$. 
The periapsis distance of the external
binary orbit was $3.3\times r_{{\rm bt}}$, 
and we fixed the initial inclination
of the inner orbits with respect to the external orbit around the
SMBH at $I=88^{\circ}$.  
We then varied $e_2$ and $a_1$ in order to 
study the dependence of the merger timescale on these latter quantities.
The choice of a large inclination was also
motivated by the short merger timescale of such systems allowing
for an efficient computation of their dynamical evolution. 

Figure~\ref{example1}
gives the time evolution of semi-major axis, eccentricity and gravitational
wave frequency~($f_{{\rm GW}}$) of the inner binaries. The peak
gravitational wave frequency was approximated by \citep{wen03}: \begin{equation}
f_{{\rm GW}}=\frac{\sqrt{GM_{b}}}{\pi}\frac{1}{\left[a_{1}(1-e_{1}^{2})\right]^{1.5}}(1+e_{1})^{1.1954}~.\end{equation}
with semi-major axis and eccentricity obtained from the relative
radial expressions given in Equations~($3.6~a$) and ($3.6~b$) of
\citet{dam+85}. 

The oscillations in the binary inner eccentricity
and gravitational wave frequency induced by the SMBH perturbations
speed up the merger time by many orders of magnitude. For example,
the nominal gravitational wave merger timescale for the cases shown
in the middle panels of the figure without SMBH is $\sim10^{10}~$yr,
while it is just few years in our simulations.
The oscillations in eccentricity and inclination observed in the computed
orbits are reminiscent of the {}``Kozai cycles'' generally discussed
in the context of hierarchical triple systems \citep[e.g.][]{1968H,WM:07,nao+11,PM:11}.
However, the distance of closest approach to the SMBH for the orbits
of Figure~\ref{example1} is too short (about 3 times the binary
tidal disruption radius) for the system to be treated exactly as
hierarchical. In other words, the strong interaction at periapsis causes high-order
terms (beyond the octupole-order) in the equations of motion
to become important to the evolution. The amplitude of the eccentricity oscillations
changes essentially at each cycle, eventually making the
inner binary periapsis distance small enough that efficient  GW loss takes place.
Interesting, we found that the changes in eccentricity, for initially high eccentric orbits ~($e_2\gtrsim 0.5$), always occur
very rapidly  during the closest approach of the binary to the SMBH rather than in
a smooth-like way along the entire orbit as it would be the case if 
we had used the secular perturbation equations 
instead of direct $N$-body integrations.
This behavior  is evident in Figure~\ref{example1}, but see also
Figure~7 in \citet{ant+10}. 

We note also that in some cases (e.g., upper left panel of Figure~\ref{example1}) it is possible that the
binary first enters a regime in which GW starts to dominate the evolution,
suddenly causing a drop of the binary semi-major axis; during this
phase the binary is then brought back to a small-eccentricity configuration
where GW loss becomes again negligible with the subsequent evolution
occurring at approximately constant semi-major axis. 

Figure~\ref{example2} displays the evolution of a stellar black hole binary with $a_{1}=0.1~$AU,
inclination $I=88^{\circ}$ and $g_{1}=0^{\circ}$. The binary is
on a circular orbit of radius $20~$AU from the central black hole.
During the oscillations in eccentricities, at the maximum of a Kozai cycle, the binary 
gravitational wave frequency is such that system  could be a detectable source for LISA-like gravitational
wave detectors. The signal from such sources in a LISA-like detector would result in repeated pulses emitted during periapsis approach.
The system merges in only $\sim20~$yr, after many eccentricity oscillations
have occurred and enters the LIGO observational window~($\gtrsim10$Hz)
with a very large eccentricity ($e_{1}\approx0.99995$) before the
fast orbital circularization due to GW loss begins. \citet{wen03}
showed that, in the contest of stellar black hole triples that might
form in the core of globular clusters, for some particular configurations
it is possible that the inner binary GW frequency enters the LIGO
observational band before the maximum eccentricity expected for the
Kozai oscillations is reached. In these cases the eccentricities can
remain significantly large ($\sim0.9$) at $f_{{\rm GW}}\sim10$Hz.
We stress that the system shown in Figure~\ref{example2} is qualitatively
different from those discussed in this previous work: only after multiple
periodic oscillations in the orbital elements have already occurred,
the binary reaches very high eccentricities and enter the LIGO frequency
band. Such a situation occurs before gravitational radiation  becomes important,
allowing the eccentricity to be so large at such high frequencies.

\subsection{Approximate method}

\label{sub:approximate-method}

Given the significant computational resources per run made necessary
by the high accuracy of the $N$-body integrations presented in the
previous section, no attempt was made to systematically explore the
large parameter space of the three body interactions. In order to perform
such broader  exploration we use in what follows an approximate method
which is detailed below.

If the external orbit of the inner binary is wide enough, i.e., the
gravitational perturbations from the SMBH on the binary are weak,
the system can be regarded as a hierarchical triple and it is possible
to treat the dynamics of the entire system as if it consists of an
inner binary of point masses $m_{0}$ and $m_{1}$ and an external
binary of masses $M_{b}=m_{0}+m_{1}$ and $M_{\bullet}$. We 
define the angular momenta $G_{1}$ and $G_{2}$ of the inner and
outer binary and total angular momentum ${\bf H}={\bf G}_{1}+{\bf G}_{2}$:
\begin{equation}
G_{1}=m_{0}m_{1}\left[\frac{Ga_{1}\left(1-e_{1}^{2}\right)}{M_{b}}\right]^{1/2}~,\end{equation}
and \begin{equation}
G_{2}=M_{b}~M_{\bullet}\left[\frac{Ga_{2}\left(1-e_{2}^{2}\right)}{M_{b}+M_{\bullet}}\right]^{1/2}~.\end{equation}

The quadrupole-level secular perturbation equations describing the
evolution of the inner binary orbital elements are \citep{for+00}
: \begin{equation}
\frac{da_{1}}{dt}=\frac{-64G^{3}m_{0}m_{1}M_{b}}{5c^{5}a_{1}^{3}
(1-e_{1}^{2})^{7/2}}\left(1+\frac{73}{24}e_{1}^{2}+\frac{37}{96}e_{1}^{4}\right)~,\label{eqm1}\end{equation}
\begin{eqnarray}
\frac{de_{1}}{dt} & = & 30K\frac{e_{1}(1-e_{1}^{2})}{G_{1}}\left(1-{\rm cos}^{2}~I\right){\rm sin}~2g_{1}\nonumber \\
&  & -\frac{304G^{3}m_{0}m_{1}M_{b}e_{1}}{15c^{5}a_{1}^{4}\left(1-e_{1}^{2}\right)^{5/2}}
\left(1+\frac{121}{304}e_{1}^{2}\right)~,\label{eq:eccen}\end{eqnarray}
\begin{eqnarray}
\frac{dg_{1}}{dt} & = & 6K\Big(\frac{1}{G_{1}}\left[4{\rm cos}^{2}~I+(5{\rm cos}~2g_{1}-1)
(1-e_{1}^{2}-{\rm cos}^{2}I)\right]\label{eq:g1}\nonumber\\
&  & +\frac{{\rm cos}~I}{G_{2}}\left[2+e_{1}^{2}(3-5{\rm cos}~2g_{1})\right]\Big) \\
&  & +\frac{3}{c^{2}a_{1}(1-e_{1}^{2})}\left[\frac{GM_{b}}{a_{1}}\right]^{3/2}~,\nonumber \end{eqnarray}
\begin{eqnarray}
\frac{dH}{dt} & = & -\frac{32G^{3}m_{0}^{2}m_{1}^{2}}{5c^{5}a_{1}^{3}(1-e_{1}^{2})^{2}}\left[\frac{GM_{b}}{a_{1}}\right]^{1/2}\label{eqm4}\\
&  & \left(1+\frac{7}{8}e_{1}^{2}\right)\frac{G_{1}+G_{2}{\rm cos}~I}{H}~,~~~~~~~~\nonumber \end{eqnarray}
where \begin{equation}
K=\frac{Gm_{0}m_{1}M_{\bullet}}{16M_{b}a_{2}(1-e_{2}^{2})^{3/2}}\left(\frac{a_{1}}{a_{2}}\right)^{2}~\end{equation}
and \begin{equation}
{\rm cos}~I=\frac{H^{2}-G_{1}^{2}-G_{2}^{2}}{2G_{1}G_{2}}~.\label{alfa}\end{equation}
Equations~(22)-(25) can be considered as describing the interaction
between two weighted ellipses instead of point masses in orbits and
are accurate as long as the energy of the system and its angular momentum
are approximately conserved quantities within each Kozai-cycle.
These equations include the effects of relativistic precession and energy
loss due to gravitational wave emission and they can be integrated numerically much more rapidly than
the full equations of motion.
They are used  here to investigate
the dynamical evolution of compact binaries near massive black holes.

We set the mass of the binary components similar to that typical of
stellar black holes: $m_{0}=m_{1}=10~M_{\odot}$\citep{woo+02}.
In this case Equations~(22)-(24) are equivalent to the octupole-level
secular perturbation equations given in \citet{bla+02}. In fact, from Equation~(19) of   \citet{bla+02} 
we see that when $m_{0}=m_{1}$ the octuple terms vanish.
We also run integrations for unequal mass binaries and including octuple terms to see the effect
that higher order terms have on the binary evolution. 
These integrations gave results very similar to those obtained using the quadruple approximation.
The fact that  the octuple terms have only a small effect on the binary dynamics is easily explained if we 
 parametrize the relative contribution of the octuple level approximation using the parameter~\citep{nao+11b}:
\begin{equation}
\epsilon_M=\left(\frac{m_0-m_1}{m_0+m_1}\right)\left(\frac{a_1}{a_2} \right) \frac{e_2}{1-e_2^2}~.
\end{equation}
In our systems  $a_2\gg a_1$ and even for very large eccentricities~($e_2\sim 0.9$) the contribution of the octuple terms is negligible.

 In Figure~\ref{ta1}
we give the merger time, $T_{M}$, as a function of the semi-major axis
of the outer orbit, this last quantity is given in units of the initial
binary semi-major axis ($a_{1}$). We set $e_{1}=e_{2}=0.1$, $a_{1}=(0.1,~0.05,~0.01)~$AU
and we consider several values of the inclination angle. The merger
time is compared to the evaporation time ($T_{{\rm EV}}$) of the
binary systems which we computed using Equation~(\ref{eq:TEV}).
The time for evaporation in a shallow density profile is greatly increased
with respect to the cusp model and it is independent on radius since,
close to the SMBH, both $\rho$ and $\sigma$ in Equation~(\ref{eq:TEV})
go like~$\sim r^{-0.5}$~.

Highly inclined binaries ($I\gtrsim70^{\circ}$) that lie close to
the central black hole inspiral and coalesce over timescales much
shorter than a Hubble time and typically shorter than their evaporation
timescale on such environments. At large distances from the SMBH relativistic
precession becomes efficient at suppressing Kozai cycles and the merger
time approaches its value when the binary evolves in isolation. Notice
that, since the perturber in our simulations is much more massive
than the inner binary, there is essentially no difference between
retrograde and prograde orbits with the minimum time for merger occurring
at $I=90^{\circ}$. Filled black circles in the figure were obtained
by direct integrations of the three body system carried out using
the \texttt{ARCHAIN} integrator. The agreement between direct integrations
and the results obtained by using the secular perturbation equations
is indeed very good.

Dashed lines in Figure~\ref{ta1} give the time required for the
stellar black holes to spiral into the central black hole. This timescale was approximated by considering the compact
binary as a single point mass and by using \citet{pet64} formula:
\begin{eqnarray}
T_{0}{_{,\bullet}} & \simeq & 1.8\times10^{12}{\rm yr}\left(\frac{20M_{\odot}}{M_{b}}\right)
\left(\frac{4\times10^{6}M_{\odot}}{M_{\bullet}}\right)^{2}\label{eq: tmer}\\
&  & \times\left(\frac{a_{2}}{{\rm mpc}}\right)^{4}(1-e_{2}^{2})^{7/2}~.\nonumber \end{eqnarray}
For small inclinations~($\lesssim70^{\circ}$), within $a_{2}/a_{1}\lesssim700$, the merger
time of the binary into the SMBH can be shorter than the time required
for the stellar black holes to merge due to gravitational wave loss.
However, the compact binary will be broken apart by tidal forces if
its center of mass approaches the SMBH within the tidal break-up radius,
$r_{\mathrm{bt}}$, which is larger than the Schwarzschild radius
of the central black hole: $r_{\mathrm{sc}}\approx0.04~{\rm AU}$.
If the compact binary does not merge due to the perturbations by the
SMBH, it will be disrupted before the two stellar black holes separately
inspiral and merge with the central object. In this case, each of
the two inspiral events cannot be treated as an isolated two-body
problem and the mutual perturbations between the stellar black holes
could produce a gravitational wave signal in a LISA-like detector
which would differ significantly from the idealized waveform obtained
in the case of an \emph{isolated} extreme-mass ratio inspiral \citep{ama+12}.

It is  interesting to explore the effect of changing outer
and inner orbital eccentricities on the binary merger time. In the
upper panel of Figure~\ref{ta} each curve corresponds to a fixed
periapsis distance and inclination $I=85^{\circ}$. As expected from
Equation~(\ref{kp}), larger external eccentricities lead initially
to faster mergers, but, as the orbital period of the external orbit
becomes comparable to the timescale for relativistic precession in
the inner binary, the merger time starts to increase rapidly with
the orbital radius, or apoapsis, of the external orbit. For $a_{1}=0.1~$AU,
setting $r_{{\rm per}}\approx20~$AU, the binary merger time is $10^{-3}\times T_{0}$
when $r_{{\rm apo}}\approx0.01~$pc. This is also an estimate of the
limiting galactocentric radius within which we expect Kozai oscillations
to significantly affect the dynamical evolution of the black hole
binaries. The lower panel of the same figure gives $T_{{\rm M}}$
as a function of $e_{1}$ for $I=85^{\circ}$, $a_{1}=0.05~$AU and
$a_{2}=130,~100$ and $70~$AU. Note that,  $T_{0}$ is also a function
of $e_{1}$, for eccentricity larger than about $e_{1}\gtrsim0.7$
the effect of the Kozai resonance on the merger time becomes progressively
less important since $e_{1}$ approaches (or exceeds) the value of
the maximum eccentricity that would be attained by the binary via
Kozai processes, which is mainly fixed by the initial inclination~($e_{\rm max}=\sqrt{1-5/3 {\rm cos^2}I}$,
neglecting  additional sources of apsidal precession).

\subsection{Merger fraction and eccentricity distribution}
\label{sec:merger_frac}

The number of GW sources from binary mergers in galactic nuclei $N_{\rm GW}$,
is the number of binaries residing in some region close to the SMBH
where they can efficiently merge $(r<R_{merger})$:\begin{eqnarray}
N_{\rm GW} &=& N_{\star}(r<R_{merge})\times f_{bin}\nonumber \\
&\times& f_{cbin}(r<a_{merge})f_{merge}(i>|90-i_{merge}|)\label{eq:N_merge}~~\end{eqnarray}
where $N_{\star}(r<R_{merge})$ is the number of stars in the enclosed
region, $f_{bin}$ is the binary fraction for the given stellar type
discussed earlier, $f_{cbin}$ is the fraction of close enough binaries
$(a<a_{merge})$ that survive in this region and $f_{merge}$ is the
fraction of binaries for which Kozai evolution lead to merger. In
the following we estimate the rate of binary mergers as well as the
properties of the coalescing binaries, which will determine their
GW signal characteristics. In order to do so we use a Monte-Carlo
approach which accounts for all the relevant parameters in Equation~(\ref{eq:N_merge}),
through random sampling of binaries with the appropriate properties.
Binary properties are sampled from the large phase-space of the various
binary properties, which we describe below. Through the random sampling,
we obtain a large ensemble of binaries and use the approximate method
described in section \ref{sub:approximate-method}, to evolve each
of the binaries and check whether it could coalesce before it evaporates
due to encounters with other stars. For those binaries that coalesce
we save the binary eccentricity as it enters the LIGO band; these
are then used to determine the overall eccentricity distribution of
the GW sources. We first determine the relative fractions of merging
binaries arising from different separations around the SMBH (see Table
\ref{Tab:rates}), and then normalize the overall rates by the expected
number of binaries residing in that region (which depends on the total
number of stars multiplied by the binary fraction). In the following,
we describe the method in details. 

We considered the idealized case of binaries with equal components
of masses $10~M_{\odot}$ or $2~M_{\odot}$ corresponding to BH-BH
and NS-NS binaries respectively. Given the uncertainty in the spatial
distribution and properties of compact binaries in galactic nuclei
the results presented in this section constitute a set of baselines
for making predictions about the expected gravitational wave signal
produced by such systems.

We sampled the binaries external orbits from the distribution of orbital
elements: \begin{equation}
N(a,e^{2})dade^{2}=N_{0}a^{2-\beta}dade^{2},\label{doe}\end{equation}
which generates steady-state phase-space distributions for an isotropic
density cusp near a dominating point mass potential. We first used the density model of Equation~(\ref{den})
with $\gamma=0.5,\,1.8$ to determine the evaporation time scale of
the binary systems and we assumed that they follow the same density distribution
of background perturbers, solar mass stars in this case. In a further
set of integrations, the evaporation time-scale was computed from
the combined density profile of stars and BHs corresponding to a mass
segregated cusp near SgrA{*} \citep{hop+06a}: \begin{equation}
\rho(r)=\rho_{{\rm \star}}(r)+\rho_{{\rm BH}}\left(\frac{r}{1{\rm pc}}\right)^{-2}~,\label{den-ms}\end{equation}
with $\rho_{{\rm BH}}=1\times10^{4}M_{\odot}{\rm pc^{-3}}$. The
functional form of the stellar density profile, $\rho_{\star}(r)$,
is that of Equation~(\ref{den}) with $\gamma=1.4$. The power law
index in equation~(\ref{doe}) was set to $\beta=2$ and $\beta=1.5$
for BHs and NSs respectively. We imposed a lower limit for the periapsis
of the external orbit of $a_{2}(1-e_{2})>4r_{{\rm bt}}$ based on
the fact that at shorter distances the binary is unstable due to the strong
perturbations from the central black hole and Equations~(3)-(6) become
a poor description of the dynamical evolution of the triple system. The distribution of inner semi-major axes, $a_{1}$,
follows the binary distribution given in Figure~1. The eccentricity of the inner binaries, $e_{1}$, were instead sampled from a thermal distribution,
i.e. $N(<e_{1})\propto e_{1}^{2}$. We adopted a uniform distribution
in ${\rm cos}(I)$ and a random distribution in $g_{1}$.

An approximate estimate of the merger time was obtained with the technique
described in~\citet{wen03} and summarized below. Including the contribution
of relativistic precession, but in the absence of gravitational radiation,
the maximum eccentricity in a Kozai cycle can be estimated by considering
that the Hamiltonian of the system, in its quadrupole form, is a conserved
quantity. The total Hamiltonian is $\mathcal{H}=kW$ where $k=3Gm_{0}m_{1}m_{2}a_{1}^2/8M_{1}a_{2}(1-e_{2}^2)^{1/2}$
and \citep{mil+02e} \begin{eqnarray}
W(g_{1},e_{1}) & = & -2(1-e_{1})+(1-e_{1}){\rm cos^2}I\nonumber \\
&  & +5e_{1}{\rm sin}2g_{1}({\rm cos^2}I-1)+\frac{4}{\sqrt{1-e_{1}}}\frac{M_{b}}{M_{\bullet}}~~~~~~~~~~\\
&  & \times\left(\frac{a_{2}\sqrt{1-e_{2}^2}}{a_1}\right)^3\left(\frac{2GM_{b}}{a_{1}c^2}\right)~,\nonumber \end{eqnarray}
with the last term accounting for relativistic precession. The system
starts from initial eccentricity $e_{1}$, argument of periapsis $g_{1}$
and mutual inclination $I$ and evolves to a maximum eccentricity
$e_{{\rm max}}$ and critical $g_{{\rm crit}}$ and $I_{{\rm crit}}$.
These latter quantities can be related to each other setting $de_{1}/dt=0$
in equation~(\ref{eq:eccen}), this gives: \begin{eqnarray}
{\rm sin}~2g_{1,{\rm crit}} & = & \frac{152G^{3}m_{0}m_{1}M_{b}}{c^{5}a_{1}^4
\left(1-e_{{\rm max}}^2\right)^{7/2}}\left(1+\frac{121}{304}e_{{\rm max}}^2\right)~\nonumber \\
&  & \times\frac{G_1}{K\left(1-{\rm cos}^2~I_{{\rm crit}}\right)}.\end{eqnarray}
Using equation~(\ref{alfa}) to relate $I_{{\rm crit}}$ to $e_{{\rm max}}$,
and assuming that $a_{1}$, $a_{2}$ and $e_{2}$ are constant quantities,
one obtains $e_{{\rm max}}$ solving the implicit equation: \begin{eqnarray}
W(g_{1},e_{1})-W(g_{{\rm crit}},e_{{\rm max}})=0~.\end{eqnarray}
An approximation to the merger time is \begin{eqnarray}
T_{{\rm merge}}=T_{{\rm 0}}(a_{1},e_{{\rm max}})/\sqrt{(1-e_{{\rm max}}^2)}~.\end{eqnarray}
Given a certain distribution of orbital elements we then estimated
the number of binaries that would merge in the galactic nucleus due
to Kozai resonances induced by the central black hole.

Table~\ref{Tab:rates} gives the total fraction of merging binaries
for which the external orbits are sampled inside some fiducial galactocentric
distances. We find that for BH-BH binaries, the SMBH-Kozai induced mergers 
represent $\gtrsim 40 \%$ of the total number of mergers occurring 
at $r\lesssim0.01~$pc. The overall fraction
of mergers in models characterized by a cusp in the distribution of background perturbers~(i.e., mass-segregated and cusp models)   is slightly lower 
with respect to that of  models with a core  due to the shorter evaporation time-scale at small radii in these former systems. When we sampled
the external orbits within larger radii~($>0.1~$pc), the merger
probability increases since the evaporation time becomes large with respect
to the merger time of the binaries. However, 
at  larger radii, both evaporation processes and GR precession
become more important in quenching Kozai cycles
that contribute  only  $\sim 10\%$ to the overall merger rate.
In the table we also give the total fraction~($f'$) of binary
systems with merger time in isolation shorter than their evaporation
time-scale. These systems would merge even with no help from the Kozai
mechanism. Comparing these numbers with those corresponding to mergers
induced by the gravitational perturbations from the SMBH, we see that 
that the process described in this paper could contribute only a small fraction of LIGO sources in galactic nuclei, 
nevertheless, these are qualitatively different  than regular mergers, as they contribute high eccentricity GW sources as shown below. 

The total numbers of
GW sources from compact binary mergers in  Milky-Way like nuclei
can now be inferred  by multiplying the fractions we obtained with
the total number of binaries in such regions, i.e. the fraction of
compact objects binaries in the stellar population. For the cusp and
core models we assume the fraction of COs resemble their population
in the field. Following \citealp{hop+06b} we take the fraction of
BHs and NSs to be 0.001 and 0.01 of the total stellar population,
respectively (in the absence of mass segregation). For the mass segregated
models, we follow the distributions obtained by \citet{hop+06b} (this
time after mass segregation occurred); i.e. we take $n_{BH}(r)\propto r^{-2}$
and $n_{NS}(r)\propto r^{-1.5}$ radial distributions, with the total
numbers normalized such that 1800 BHs and 374 NSs reside inside 0.1
pc . We assume the binary fractions follow our discussion in section
\ref{sec:Binaries} (0.1 and 0.07 for BHs and NSs, respectively).
Taking these numbers we calculate the total numbers of merging compact
binaries in a given Milky-Way like galactic nucleus (see Table 1). 

\begin{figure}
\centering
\includegraphics[width=2.53in,angle=270.]{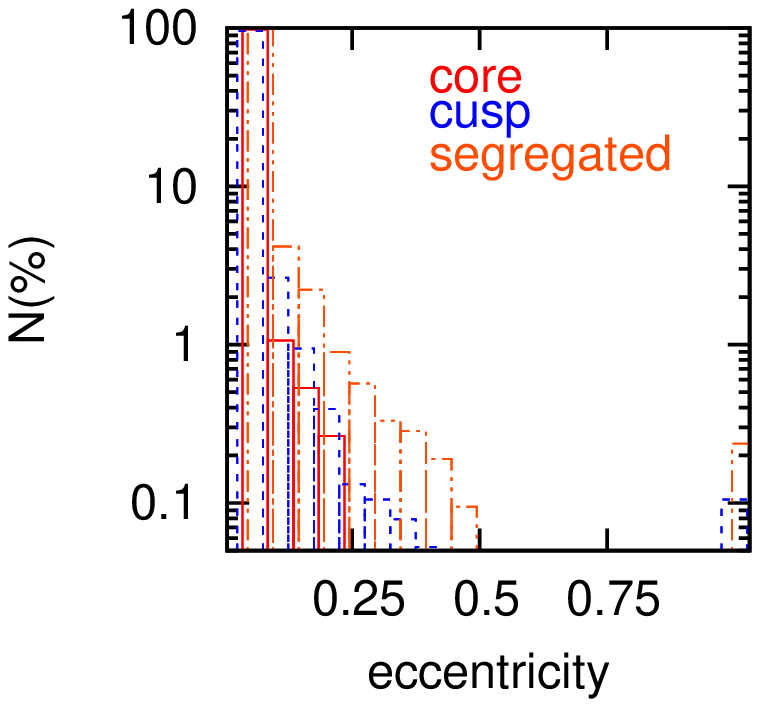}
\includegraphics[width=2.53in,angle=270.]{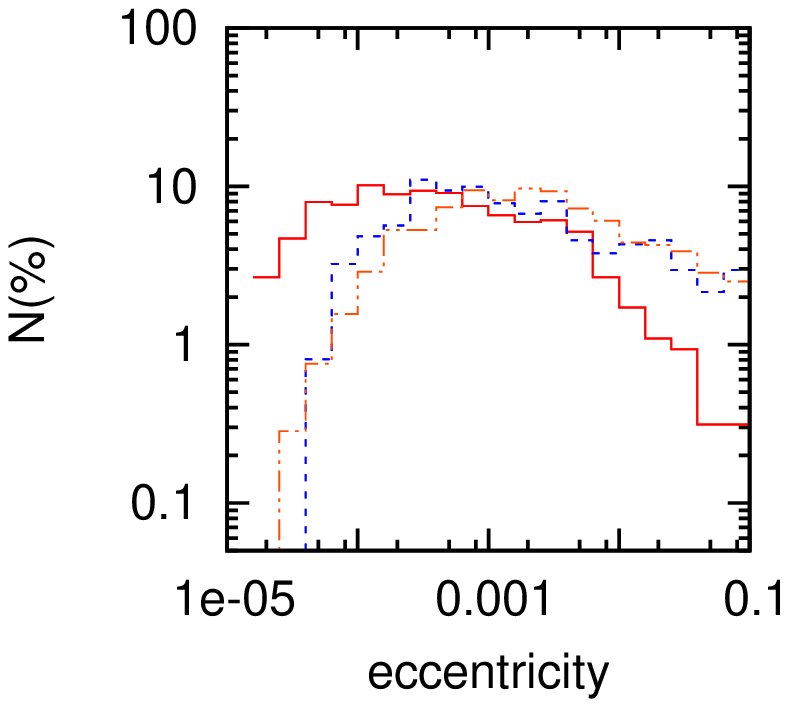}
 \caption{Distribution of eccentricity at $f_{{\rm GW}}=10~$Hz~
for BH-BH mergers driven by Kozai effects in a Milky
Way-like nucleus. The external binary orbits were generated from the
distribution of orbital elements given in equation~(\ref{doe}) where
we adopted $\beta=0.5$~(core; continue-line histograms), 1.8~(cusp;
dashed-line histograms) and 2~(relaxed cusp; dash-dotted histograms).
The distribution of background perturbers from
which we computed the binaries evaporation times was either the broken
power-law model of equation~(\ref{den}) with $\gamma=\beta$ for
the {}``core'' and the {}``cusp'' models or the density model
of equation~(\ref{den-ms}) for the mass-segregated model. The distribution
of semi-major axes, $a_{1}$, followed that given in Figure~1, while
the internal eccentricities $e_{1}$ were sampled from a thermal distribution~(i.e.,~$N(<e_{1})\propto e_{1}^{2}$). 
Lower panel gives the details of the eccentricity distribution for $e_1<0.1$. For the cusp and the mass-segregated models
most of the mergers occur at $e_1>10^{-4}$.}
\label{fig:ist} 
\end{figure}

Figure~\ref{fig:ist} shows the number distribution of eccentricities
at $f_{{\rm GW}}=10~$Hz~(i.e., the beginning of the LIGO band) for
merging BH-BH binaries. This plot was obtained by integrating the set of
equations~(\ref{eqm1})-(\ref{eqm4}) and sampling the initial conditions
as detailed above but considering only systems with inclinations $I\geq~80^{\circ}$.
The permitted parameter space was defined by imposing the evaporation time of the
binary to be longer than its merger time.
 As additional constrain we imposed  the merger time of the binary in isolation to be shorter
than the local evaporation time (i.e., $T_{{\rm 0}}(a_{1},e_{1})>T_{{\rm EV}}/\sqrt{1-e_2^2}$)
since these systems would rapidly merge even without undergoing 
Kozai resonance.

In all cases, at $f_{{\rm GW}}=10~$Hz about $\sim90\%$ of merging
systems have $e_{1}<0.1$. However, in the cusp and mass-segregated
distributions a small fraction~(about $\sim 0.1\%$) of merging
binaries have very large eccentricities $e_{1}\sim 1$ at the moment they enter
the LIGO band. These are cases in which the gravitational radiation
effect dominates  the evolution within one Kozai cycle \citep[see e.g. Figure~8 of][]{wen03}.
About $0.5\%$ of the mergers have eccentricities larger than $0.5$ at $10~$Hz.
Systems that reach such high eccentricities at the first Kozai maximum
will be very short lived, which decreases the probability that
they can be observed. On the other hand, eccentric mergers might be
detectable  
to larger distances~\citep[$\sim1~$Gpc for advanced-LIGO and for $10~M_{\odot}$ black holes, e.g.,]{O:09}
and greater black hole masses than circular mergers. Finally, we note  that the fraction of eccentric mergers found here is a conservative 
estimate due to the  lower limit that 
we imposed for the periapsis of the external orbit (i.e., $a_{2}(1-e_{2})>4r_{{\rm bt}}$). 
 In fact, we  expect that for smaller periapsis,  the binary could
  evolve through a non-secular evolution that could lead to
very large eccentricities in only a few years~(e.g., Figure~\ref{example2}).

\subsection{GW rates from coalescing binaries in galactic nuclei}

\label{sub:GW-rates}

In the previous section we calculated the total number of merging
compact binaries, for a given galactic nucleus. The total GW rated
from such mergers depends on the rate at which the binary progenitors
of such coalescing binaries are replenished. The GW rate form coalescing
binaries in a galactic nucleus is therefore $\Gamma_{\rm GW}=f_{\rm GW}\times\Gamma_{rep}$,
where $\Gamma_{rep}$ is the rate of binaries replenished into the
GC timescale (if no replenishment occurs, we take the replenishment
rate to be the total number of primordial binaries divided by the
age of the Universe). We estimate the replenishment rate assuming it is 
dominated by diffusional replenishment (see below), but we discuss various other 
mechanisms which may contribute to increase the replenishment rate, beyond our basic estimates. 

\textbf{Diffusional replenishment:} The simplest form of replenishment
is through diffusional migration of binaries outside the region of
influence of the SMBH in the central region. The timescale for such
processes (e.g. as studied by \citealp{hop09}) is the relaxation
timescale at which the system approaches a steady state with no memory
of the initial conditions. The replenishment rate assuming such a
steady state is of the order of the number of CO binaries, given a
nucleus model, divided by the relaxation time, and we get $\Gamma_{\rm GW}=N_{\rm GW}/\min(\tau_{rep},\,\tau_{H})$.
The calculated GW rates, given these values are shown in Table~1.
The relaxation timescales are taken from Fig. \ref{fig:times}; these
are approximately constant for the cusp model throughout the central
pc ($\tau_{rep}(r)\sim Const=10^{9}$ yrs), and are longer than a
Hubble time for the core models (and we therefore take $\Gamma_{\rm GW}=N_{\rm GW}/\tau_{H}$).

\textbf{Comparison with other GW sources and rate estimates:} When compared
with other GW sources outside galactic nuclei the GW rates from the
processes discussed here are typically lower than most other estimates
discussed in \citet{aba+10}; reflecting the small number of compact objects residing in a 
galactic nucleus compared with the total number of such objects in a galaxy. 
The maximum BH-BH coalescence rate we find (for the segregated cusp model), provides a 
coalescence rate which is five times larger than the low estimates in \citet{aba+10}, 
10 \% of the realistic rate and 0.1 \% of the high/optimistic rate estimate. The contribution to 
NS-NS coalescence rate is small in all cases; the highest rate we find (for the cusp model) is 
10, 1 and 0.1 \% of the low, realistic and high estimates in \citet{aba+10}, respectively. 
The overall contribution of GW sources from galactic nuclei is therefore likely to be small, however, 
the eccentric signature of a fraction of these sources is qualitatively different than regular GW sources.

\begin{table*}
\centering
\begin{tabular}{llllll}
\hline 
& $f(\%);~f'(\%)$  & N$_{GW}$ & $\Gamma_{GW}$ & $f(\%);~f'(\%)$  & $f(\%);~f'(\%)$ \tabularnewline
& $r<0.5$ pc  &  & (Myr$^{-1}$) & $r<0.1$ pc  & $r<0.01$ pc \tabularnewline
\hline 
BH-BH &  &  &  &  & \tabularnewline
\hline 
core  & 8.9; 8.1  & $2.2$ & $1.7\times10^{-4}$ & 9.46;~8.15  & 14.0;~8.18 \tabularnewline
cusp  & 5.9; 5.2  & $3.6$ & $3.6\times10^{-3}$ & 4.01;~3.06 & 2.81;~1.49 \tabularnewline
segregated  & 5.5; 4.8  & $49.5$ & $4.8\times10^{-2}$ & 3.01;~2.61  & 2.03;~1.16 \tabularnewline
\hline 
NS-NS & \multicolumn{1}{c}{} &  & \tabularnewline
\hline 
core  & 32.9; 31.5  & 54.8 & $3.9\times10^{-4}$ & 33.1;~32.2  & 39.5; 32.2 \tabularnewline
cusp  & 26.1; 24.6  & 124 & $1.2\times10^{-1}$ & 17.7;~ 17.2  & 9.01; 6.36 \tabularnewline
segregated  & 43.5; 41.7  & 1.3 & $1.3\times10^{-3}$ & 28.6;~ 27.7 & 10.8; 9.55\tabularnewline
\hline &  &  &  &  & \tabularnewline
\end{tabular}

\caption{\label{Tab:rates} GW rate estimates and the total fraction of BH-BH
and NS-NS binaries merging due to Kozai resonances in a galactic nucleus~($f$),
and fraction of mergers that would occur in a galactic nucleus with
no help from the central SMBH~($f'$). All the GW rate estimates
are per Milky-Way like galaxy. }

\end{table*}

\textbf{Other replenishment processes: }Various other processes may
affect the replenishment rate of compact binaries and/or their progenitors
in the GC. In the following we qualitatively discuss such processes,
but we do not account for them in our rates estimates (beside star
formation, effectively integrated in into the assumptions used in
the observations based estimates), as their complete analysis is beyond
the scope of this paper.

\emph{Triple disruption}\textbf{:} \citeauthor{per09b} (2009; see
also Ginsburg and Perets 2011) suggested that disruption of triple
stars could leave behind a binary in a close orbit around the SMBH;
this could serve as a continuous source of replenished binaries close
to the SMBH. This study estimated a triple disruption rate which is
$\sim f_{triple}=3-5$ \% of the binary disruption rate of massive stars
(OB stars; we do not consider disruption of triple compact objects
as these are likely to be rare). In half of these cases a binary is
captured around the SMBH, where its companion is ejected. If the S-stars
observed close to the SMBH in the GC arise from a binary disruptions
and/or similarly the observed population of HVSs arise from such processes,
then these could be used to calibrate the binary and hence triple
disruption rates. Capture rate calculations by \citeauthor{per+07}\citep[2007; see also][]{mad+09,per+09b}, suggest that the binary
disruption scenario (induced by massive perturbers scattering) could
be consistent with the observed number of the young B-stars in the
GC (and similar processes are likely to be at work in other galaxies;
\citealp{per+08d}). If this is the case then binaries are continuously
replenished into close regions around the SMBH (and into eccentric
orbits around the SMBH), and our use of the observed number of B-stars
could be justified, with some important differences. Only $\sim1.5-3$
\% of the observed B-stars would be the result of triple disruption
and would actually be binaries. On the other hand the captured binaries
are likely to have smaller separations than typical binaries in the
field, which are more likely to form compact binary progenitors that
would efficiently merge. Taken together, it is likely that such a
replenishment mechanism would produce a GW rate which is 10-20
times lower then the high estimate for the NS stars. This scenario,
however, is not likely to produce progenitors massive enough to form SBHS, such as the observed young O/WR stars in
the GC (which reside in a disk like configuration and have low eccentricities).
GW rate from BHs coalescence are therefore not likely to gain much from this
process. 

\emph{In-situ star-formation:}\textbf{ }As mentioned above, our GC
contains a large population of young massive O-stars, many of which
reside in a stellar disk. These stars were most likely to form in-situ
following the fragmentation of a gaseous disk formed from an in-falling
gaseous clump \citep[][and references therein]{gen+10}. Such stars
formation burst may occur continuously throughout the evolution of
the stellar cusp around the SMBH. As mentioned earlier eclipsing and
close binaries are observed among the GC young stars, suggesting star
formation as an additional process which can repopulate the binary
population of the GC. Continuous formation of binaries in the GC is
part of the assumptions made in our high GW rate estimate. 

The other assumptions in the high estimate are related to the binary
population properties and their distribution around the SMBH. Although
some pioneering work have been done on the properties of binaries
formed in the GC \citep{ale+08}, these are generally unknown. Another
important issue is whether the orbits of binaries formed in a disk
could later dynamically evolve to achieve close approach to the SMBH,
where efficient secular evolution as studied here can occur. Although
the O-stars observed in the disks are not observed closer than 0.05
pc to the SMBH, various mechanisms were suggested for their inward
migration closer to the SMBH. \citet{bar+11} discussed the disk migration
of binaries through similar migration scenario of planets in gaseous
disks; they show that the binaries do not only migrate but also shrink
their inner orbit and possibly merge before they finish their migration.
Madigan et al. \citeyearpar{mad+09,mad+11} proposed that binaries
(and single stars) could be excited to high eccentricities through
an eccentric disk instability scenario, and their periapsis approach
could become small enough for the binaries to be disrupted. Both these
mechanisms suggest the existence of binaries at close approach to
the SMBH. In the disk migration scenario the binaries' orbit around
the SMBH are likely to be relatively circular, and the binaries loose
their orbital energy and migrate on a short timescale of the order
$T_{mig}\sim10^{5}$ yrs; it is not clear, however, whether the gaseous
disk enabling the migration would extend close to the SMBH, and/or
allow for migration in the innermost regions%
\footnote{It is interesting to note, however, that compact binary mergers in
a gaseous disk could provide additional and different channel for
the formation of GW sources, if compact binaries exist in the gaseous
disk, as could be the case of long lived AGN disks. We leave further
study of these issues to future work. %
}. In the eccentric disk instability binaries remain at large semi-major axis orbits
around the SMBH, but gradually increase their eccentricity, and therefore
come closer to the SMBH only at periapsis. The timescale for the
close approach is relatively short, $T_{ecc}\sim few\times10^{5}$,
and binaries may be left at eccentric orbits at the end of this process.

\emph{Cluster infall:}\textbf{ }Nuclear stellar clusters may result
from the continuous infall of globular clusters that disperse close
or in the close vicinity of the SMBH (see \citealp{ant+11} and references
therein). In particular such clusters may harbor an inner core cluster
of BHs that formed during the cluster evolution. If these BHs are
retained in the cluster this mechanism may contribute to the BHs population
in the GC. Further work needs to be done to explore the implications
of this scenario to the replenishment of binaries in the GC.\textbf{ }

\emph{Resonant relaxation:} In our discussion on replenishment  processes above
we considered  the replenishment of binaries through non-resonant relaxation
processes in which stars diffuse through random encounters with other
stars. {Resonant} relaxation could change the eccentricities
and inclination of stars much faster than regular relaxation. One
may therefore consider the timescale $\tau_{rep}$ for diffusional
replenishment to be the scalar resonant relaxation timescale $\tau_{rep}=T_{\rm RR}$,
which could be much shorter than the non-resonant relaxation timescale
assumed in the low estimate (see Fig. \ref{fig:times}); this could
raise the low estimates by 1-2 orders of magnitude. For the high estimate,
resonant relaxation may support the assumption that the distribution
of stars rapidly achieve a steady state thermal distribution of eccentricities,
as assumed in our models.  

The role of the much faster VRR, which changes
the inclination of an orbit around the SMBH is less clear. It depends
on the  dynamical effect on the inner binary inclination in respect
to its orbit around the SMBH. We may consider two extreme possibilities.
If the binary orbit conserves its mutual inclination, i.e. acts as
a stiff object, than VRR would not affect the
secular evolution of the inner binary. However, if the binary orbit
plays the role of a gyroscope, an initial co-planer configuration
(e.g. zero mutual inclination), in which Kozai evolution is ineffective,
could be transformed into a high mutual inclination configuration
at which Kozai evolution plays an important role. To further complicate
the problem, the timescales for VRR could be
short enough as to be comparable to the Kozai timescale, in which
case the coupling of both processes should be considered. All of the
above mentioned issues for VRR of binaries
require a dedicated study which is beyond the scope of this paper.

\section{Combined Kozai-cycles and tidal friction evolution of stellar binaries
near SMBHs }

\label{sec:KCTF}

In the previous sections we discussed the secular evolution of CO
binaries near a SMBH. A similar secular evolution could be important
for stellar binaries and/or MS star-CO binaries. In these cases GW emission
which served to dissipate energy and induce the inspiral of the CO
binaries is not likely to play an important role. However, binary
stars can dissipate energy through tidal friction. The secular evolution
of the triple system, SMBH-stellar binary, can therefore lead to a
similar outcomes to those studied in the context of stellar triples
(see also \citealp{per09b}). 

 \citet{1979MS} first suggested that
Kozai cycles and tidal friction in triples can induce their inspiral
and the formation of compact binaries. Later studies by \citet{2006EK}
suggested that short period binaries typically form through this process
(which could be even more effective than previously though when accounting
for octupole approximation; \citealp{nao+11b}).
\citet{per+09c} suggested that such process can induce mergers of
binary stars and the formation of blue stragglers. 
Such mergers could be therefore a possible source of young stars at the Galactic center~\citep{per+09c,ant+11}.
Similarly other types of close binaries and their products could be formed in a similar
way. We may therefore expect that the perturbative effect of SMBHs
on binaries provide an efficient channel for the formation of close
binaries and their products, such as X-ray binaries as well supernovae
and gamma ray bursts (following the merger of two WDs or NS, or through
the formation of a WD accreting binary; see also \citealp{tho+11}).
 We note, that the binary fractions of MS stars are generally higher than those of COs, as
discussed in section 2, and tidal fraction is more efficient in inducing binary shrinkage/coalescence than is GW emission. 
Taken together, we therefore expect the rate of shrinkage/merger of MS stars to higher than our estimtes of CO binary mergers.   
Since this study focuses on secular evolution of CO binaries, we postpone futher quantitative discussion of the evolution of binaries with MS star components 
to another study.

\section{Conclusion}
\label{sec:summary}

In this study we explored the secular evolution of compact binaries
orbiting a massive black hole, and considered the effects of the stellar
environment of the binaries. We have shown that the SMBH can drive the
binary (inner) orbit into high eccentricities, at which point the BHs
can efficiently coalesce through GW emission, at times shorter that
the evaporation times of the binaries in the hostile stellar environment
of the SMBH. Such coalescing binaries can have non-negligible eccentricity
when they enter the LIGO band, and therefore have a unique observable
signature. In addition we found that binaries at very close orbits
to the SMBH can interact with it on dynamical timescale and present
peculiar evolution leading to several GW pulses over
a short timescale. The latter sources are likely to be rare, and not
be detected with current or planned instruments. Finally, we note
that similar processes can be important for the formation and merger
of close stellar binaries, in which cases the GW dissipation leading
to a the binary inspiral is replaced with tidal friction.

\bigskip
We are extremely grateful to S. Mikkola who wrote the ARCHAIN integrator. We thank S.~Prodan for
a careful reading of an earlier version of this paper.
We thank K.~Cannon,  C. Hopman, B. Kocsis, D. Merritt, A.~Mrou\'{e}, N. Murray, Harald Pfeiffer and Y.~Wu for useful comments.
FA acknowledges partial support from  Harvard/CFA Predoctoral Fellowship  during the Summer 2011.

\bibliographystyle{apj}

\begin{thebibliography}{80}
\expandafter\ifx\csname natexlab\endcsname\relax\def\natexlab#1{#1}\fi

\bibitem[{{Abadie} {et~al.}(2010){Abadie}, {Abbott}, {Abbott}, {Abernathy},
 {Accadia}, {Acernese}, {Adams}, {Adhikari}, {Ajith}, {Allen}, \&
 et~al.}]{aba+10}
{Abadie}, J., {Abbott}, B.~P., {Abbott}, R., {Abernathy}, M., {Accadia}, T.,
 {Acernese}, F., {Adams}, C., {Adhikari}, R., {Ajith}, P., {Allen}, B., \&
 et~al. 2010, Classical and Quantum Gravity, 27, 173001

\bibitem[{{Alexander} {et~al.}(2008){Alexander}, {Armitage}, \&
 {Cuadra}}]{ale+08}
{Alexander}, R.~D., {Armitage}, P.~J., \& {Cuadra}, J. 2008, \mnras, 389, 1655

\bibitem[{{Alexander}(2005)}]{ale05}
{Alexander}, T. 2005, Phys. Rep., 419, 65

\bibitem[{{Amaro-Seoane} {et~al.}(2012){Amaro-Seoane}, {Brem}, {Cuadra}, \&
 {Armitage}}]{ama+12}
{Amaro-Seoane}, P., {Brem}, P., {Cuadra}, J., \& {Armitage}, P.~J. 2012, \apjl,
 744, L20

\bibitem[{{Antonini} {et~al.}(2011b){Antonini}, {Capuzzo-Dolcetta},
 {Mastrobuono-Battisti}, \& {Merritt}}]{ant+11}
{Antonini}, F., {Capuzzo-Dolcetta}, R., {Mastrobuono-Battisti}, A., \&
 {Merritt}, D. 2011, ApJ, 750, 111

\bibitem[{{Antonini} {et~al.}(2010){Antonini}, {Faber}, {Gualandris}, \&
 {Merritt}}]{ant+10}
{Antonini}, F., {Faber}, J., {Gualandris}, A., \& {Merritt}, D. 2010, \apj,
 713, 90

\bibitem[Antonini et al.~(2011b)]{ant+11a}Antonini F., Lombardi J., \& Merritt D., 2011, ApJ, 731, 128

\bibitem[Antonini \& Merritt~(2012)]{ant+12}Antonini F., \& Merritt D., 2012, ApJ, 741, 83

\bibitem[{{Bahcall} \& {Wolf}(1977)}]{bah+77}
{Bahcall}, J.~N. \& {Wolf}, R.~A. 1977, ApJ, 216, 883

\bibitem[{{Bartko} {et~al.}(2009)}]{bar+09}
{Bartko}, H. {et~al.} 2009, ApJ, 697, 1741

\bibitem[{{Bartko} {et~al.}(2010)}]{bar+10}
---. 2010, ApJ, 708, 834

\bibitem[{{Baruteau} {et~al.}(2011){Baruteau}, {Cuadra}, \& {Lin}}]{bar+11}
{Baruteau}, C., {Cuadra}, J., \& {Lin}, D.~N.~C. 2011, \apj, 726, 28

\bibitem[{{Belczynski} {et~al.}(2004){Belczynski}, {Sadowski}, \&
 {Rasio}}]{bel+04a}
{Belczynski}, K., {Sadowski}, A., \& {Rasio}, F.~A. 2004, \apj, 611, 1068

\bibitem[{{Binney} \& {Tremaine}(1987)}]{bin+87}
{Binney}, J. \& {Tremaine}, S. 1987, Galactic Dynamics (Princeton, NJ:
 Princeton University Press)

\bibitem[{{Blaes} {et~al.}(2002){Blaes}, {Lee}, \& {Socrates}}]{bla+02}
{Blaes}, O., {Lee}, M.~H., \& {Socrates}, A. 2002, ApJ, 578, 775

\bibitem[{{Buchholz} {et~al.}(2009)}]{buc+09}
{Buchholz}, R.~M. {et~al.} 2009, \aap, 499, 483

\bibitem[{{Cuadra} {et~al.}(2008){Cuadra}, {Armitage}, \& {Alexander}}]{cua+08}
{Cuadra}, J., {Armitage}, P.~J., \& {Alexander}, R.~D. 2008, MNRAS, 388, L64

\bibitem[{{Damour} \& {Deruelle}(1985)}]{dam+85}
{Damour}, T. \& {Deruelle}, N. 1985, Ann.~Inst.~Henri Poincar{\'e}
 Phys.~Th{\'e}or., Vol.~43, No.~1, p.~107 - 132, 43, 107

\bibitem[{{DePoy} {et~al.}(2004){DePoy}, {Pepper}, {Pogge}, {Stutz},
 {Pinsonneault}, \& {Sellgren}}]{dep+04}
{DePoy}, D.~L., {Pepper}, J., {Pogge}, R.~W., {Stutz}, A., {Pinsonneault}, M.,
 \& {Sellgren}, K. 2004, \apj, 617, 1127

\bibitem[{{Do} {et~al.}(2009)}]{do+09}
{Do}, T. {et~al.} 2009, \apj, 703, 1323

\bibitem[{{Duquennoy} \& {Mayor}(1991)}]{duq+91}
{Duquennoy}, A. \& {Mayor}, M. 1991, A\&A, 248, 485

\bibitem[{{Eggleton} \& {Kiseleva-Eggleton}(2001)}]{2001EK}
{Eggleton}, P.~P. \& {Kiseleva-Eggleton}, L. 2001, \apj, 562, 1012

\bibitem[{{Eggleton} \& {Kisseleva-Eggleton}(2006)}]{2006EK}
{Eggleton}, P.~P. \& {Kisseleva-Eggleton}, L. 2006, \apss, 304, 75

\bibitem[{{Ford} {et~al.}(2000){Ford}, {Kozinsky}, \& {Rasio}}]{for+00}
{Ford}, E.~B., {Kozinsky}, B., \& {Rasio}, F.~A. 2000, \apj, 535, 385

\bibitem[{{Freitag} {et~al.}(2006){Freitag}, {Amaro-Seoane}, \&
 {Kalogera}}]{fre+06}
{Freitag}, M., {Amaro-Seoane}, P., \& {Kalogera}, V. 2006, ApJ, 649, 91

\bibitem[{{Garmany} {et~al.}(1980){Garmany}, {Conti}, \& {Massey}}]{gar+80}
{Garmany}, C.~D., {Conti}, P.~S., \& {Massey}, P. 1980, ApJ, 242, 1063

\bibitem[{{Gebhardt} {et~al.}(2000)}]{geb+00}
{Gebhardt}, K. {et~al.} 2000, ApJl, 539, L13

\bibitem[{{Gebhardt} {et~al.}(2003)}]{geb+03}
---. 2003, ApJ, 583, 92

\bibitem[{{Genzel} {et~al.}(2010){Genzel}, {Eisenhauer}, \&
 {Gillessen}}]{gen+10}
{Genzel}, R., {Eisenhauer}, F., \& {Gillessen}, S. 2010, Reviews of Modern
 Physics, 82, 3121

\bibitem[{{Ghez} {et~al.}(2008)}]{ghe+08}
{Ghez}, A.~M. {et~al.} 2008, \apj, accepted (astro-ph/0808.2870)

\bibitem[{{Gillessen} {et~al.}(2009){Gillessen}, {Eisenhauer}, {Trippe},
 {Alexander}, {Genzel}, {Martins}, \& {Ott}}]{gil+09}
{Gillessen}, S., {Eisenhauer}, F., {Trippe}, S., {Alexander}, T., {Genzel}, R.,
 {Martins}, F., \& {Ott}, T. 2009, ApJ, 692, 1075

\bibitem[{{Gould}(2011)}]{gou11}
{Gould}, A. 2011, \apjl, 729, L23

\bibitem[{{Gould} \& {Quillen}(2003)}]{gou+03}
{Gould}, A. \& {Quillen}, A.~C. 2003, ApJ, 592, 935

\bibitem[Gualandris \& Merritt~(2012)]{gm+12} Gualandris, A., \& Merritt, D. \ 2012, ApJ, 744, 74

\bibitem[{{Hansen} {et~al.}(2004){Hansen}, {Kawaler}, \& {Trimble}}]{han+04}
{Hansen}, C.~J., {Kawaler}, S.~D., \& {Trimble}, V. 2004, {Stellar interiors :
 physical principles, structure, and evolution}, ed. {Hansen, C.~J., Kawaler,
 S.~D., \& Trimble, V.}

\bibitem[{{Harrington}(1968)}]{1968H}
{Harrington}, R.~S. 1968, \aj, 73, 190

\bibitem[{{Heggie}(1975)}]{heg75}
{Heggie}, D.~C. 1975, MNRAS, 173, 729

\bibitem[{{Hills}(1988)}]{hil88}
{Hills}, J.~G. 1988, Nature, 331, 687

\bibitem[{{Hollywood} \& {Melia}(1997)}]{hol+97}
{Hollywood}, J.~M. \& {Melia}, F. 1997, ApJs, 112, 423

\bibitem[{{Hopman}(2009)}]{hop09}
{Hopman}, C. 2009, \apj, 700, 1933

\bibitem[{{Hopman} \& {Alexander}(2006{\natexlab{a}})}]{hop+06a}
{Hopman}, C. \& {Alexander}, T. 2006{\natexlab{a}}, ApJ, 645, 1152

\bibitem[{{Hopman} \& {Alexander}(2006{\natexlab{b}})}]{hop+06b}
---. 2006{\natexlab{b}}, ApJl, 645, L133

\bibitem[{{Kiseleva} {et~al.}(1998){Kiseleva}, {Eggleton}, \&
 {Mikkola}}]{kis+98}
{Kiseleva}, L.~G., {Eggleton}, P.~P., \& {Mikkola}, S. 1998, MNRAS, 300, 292

\bibitem[{{Kobulnicky} \& {Fryer}(2007)}]{kob+07}
{Kobulnicky}, H.~A. \& {Fryer}, C.~L. 2007, \apj, 670, 747

\bibitem[{{Kouwenhoven} {et~al.}(2007){Kouwenhoven}, {Brown}, {Portegies
 Zwart}, \& {Kaper}}]{kou+07}
{Kouwenhoven}, M.~B.~N., {Brown}, A.~G.~A., {Portegies Zwart}, S.~F., \&
 {Kaper}, L. 2007, \aap, 474, 77

\bibitem[{{Kozai}(1962)}]{1962K}
{Kozai}, Y. 1962, \aj, 67, 591

\bibitem[{{Lada}(2006)}]{lad06}
{Lada}, C.~J. 2006, ApJl, 640, L63

\bibitem[{{Lidov}(1962)}]{1962L}
{Lidov}, M.~L. 1962, \planss, 9, 719

\bibitem[{{Lu} {et~al.}(2009)}]{lu+09}
{Lu}, J.~R. {et~al.} 2009, ApJ, 690, 1463

\bibitem[{{Madigan} {et~al.}(2011){Madigan}, {Hopman}, \& {Levin}}]{mad+11}
{Madigan}, A.-M., {Hopman}, C., \& {Levin}, Y. 2011, \apj, 738, 99

\bibitem[{{Madigan} {et~al.}(2009){Madigan}, {Levin}, \& {Hopman}}]{mad+09}
{Madigan}, A.-M., {Levin}, Y., \& {Hopman}, C. 2009, ApJl, 697, L44

\bibitem[{{Manchester} {et~al.}(2005){Manchester}, {Hobbs}, {Teoh}, \&
 {Hobbs}}]{man+05}
{Manchester}, R.~N., {Hobbs}, G.~B., {Teoh}, A., \& {Hobbs}, M. 2005, \aj, 129,
 1993

\bibitem[{{Martins} {et~al.}(2006)}]{mar+06b}
{Martins}, F. {et~al.} 2006, \apjl, 649, L103

\bibitem[{{Mason} {et~al.}(1998){Mason}, {Gies}, {Hartkopf}, {Bagnuolo}, {ten
 Brummelaar}, \& {McAlister}}]{mas+98}
{Mason}, B.~D., {Gies}, D.~R., {Hartkopf}, W.~I., {Bagnuolo}, Jr., W.~G., {ten
 Brummelaar}, T., \& {McAlister}, H.~A. 1998, \aj, 115, 821

\bibitem[{{Mazeh} \& {Shaham}(1979)}]{1979MS}
{Mazeh}, T. \& {Shaham}, J. 1979, \aap, 77, 145

\bibitem[{{Merritt}(2010)}]{mer10}
{Merritt}, D. 2010, \apj, 718, 739

\bibitem[{{Merritt} {et~al.}(2011){Merritt}, {Alexander}, {Mikkola}, \&
 {Will}}]{mer+11}
{Merritt}, D., {Alexander}, T., {Mikkola}, S., \& {Will}, C.~M. 2011, \prd, 84,
 044024

\bibitem[{Mikkola} \& {Merritt}~(2008)]{mik+08} Mikkola, S., \& Merritt, D. 2008, AJ, 135, 2398 

\bibitem[{{Miller} \& {Hamilton}(2002)}]{mil+02e}
{Miller}, M.~C. \& {Hamilton}, D.~P. 2002, \apj, 576, 894

\bibitem[{{Miller} {et~al.}(2005)}]{mil+05}
{Miller}, M.~C. {et~al.} 2005, ApJl, 631, L117

\bibitem[{{Muno} {et~al.}(2005){Muno}, {Pfahl}, {Baganoff}, {Brandt}, {Ghez},
 {Lu}, \& {Morris}}]{mun+05}
{Muno}, M.~P., {Pfahl}, E., {Baganoff}, F.~K., {Brandt}, W.~N., {Ghez}, A.,
 {Lu}, J., \& {Morris}, M.~R. 2005, ApJl, 622, L113

\bibitem[{{Naoz} {et~al.}(2011a){Naoz}, {Farr}, {Lithwick}, {Rasio}, \&  {Teyssandier}}]{nao+11}
{Naoz}, S., {Farr}, W.~M., {Lithwick}, Y., {Rasio}, F.~A., \& {Teyssandier}, J.
\ 2011, Nature, 473, 187

\bibitem[{{Naoz} {et~al.}(2011b){Naoz}, {Farr}, {Lithwick}, {Rasio}, \&  {Teyssandier}}]{nao+11b}
{Naoz}, S., {Farr}, W.~M., {Lithwick}, Y., {Rasio}, F.~A., \& {Teyssandier}, J.
\ 2011, arXiv1108.5176

\bibitem[{{O'Leary} {et~al.}(2009){O'Leary}, {Kocsis}, \& {Loeb}}]{O:09}
{O'Leary}, R.~M., {Kocsis}, B., \& {Loeb}, A. 2009, \mnras, 395, 2127

\bibitem[{{Ott} {et~al.}(1999){Ott}, {Eckart}, \& {Genzel}}]{ott+99}
{Ott}, T., {Eckart}, A., \& {Genzel}, R. 1999, ApJ, 523, 248

\bibitem[{{Peeples} {et~al.}(2007){Peeples}, {Stanek}, \& {Depoy}}]{pee+07}
{Peeples}, M.~S., {Stanek}, K.~Z., \& {Depoy}, D.~L. 2007, Acta Astronomica,
 57, 173

\bibitem[{{Perets}(2009)}]{per09b}
{Perets}, H.~B. 2009, ApJ, 698, 1330

\bibitem[{{Perets} \& {Alexander}(2008)}]{per+08d}
{Perets}, H.~B. \& {Alexander}, T. 2008, \apj, 677, 146

\bibitem[{{Perets} \& {Fabrycky}(2009)}]{per+09c}
{Perets}, H.~B. \& {Fabrycky}, D.~C. 2009, ApJ, 697, 1048

\bibitem[{{Perets} {et~al.}(2007){Perets}, {Hopman}, \& {Alexander}}]{per+07}
{Perets}, H.~B., {Hopman}, C., \& {Alexander}, T. 2007, ApJ, 656, 709

\bibitem[{{Perets} {et~al.}(2008){Perets}, {Kupi}, \& {Alexander}}]{per+08c}
{Perets}, H.~B., {Kupi}, G., \& {Alexander}, T. 2008, in IAU Symposium, Vol.
 246, IAU Symposium, ed. {E.~Vesperini, M.~Giersz, \& A.~Sills}, 275--276

\bibitem[{{Perets} {et~al.}(2009)}]{per+09b}
{Perets}, H.~B. {et~al.} 2009, ApJ, 702, 884

\bibitem[{{Peters}(1964)}]{pet64}
{Peters}, P.~C. 1964, Physical Review, 136, 1224

\bibitem[Prodan \& Murray~(2011)]{PM:11}Prodan, S. \& Murray, N. \ 2012, ApJ, 747, 4

\bibitem[{{Rafelski} {et~al.}(2007){Rafelski}, {Ghez}, {Hornstein}, {Lu}, \&
 {Morris}}]{raf+07}
{Rafelski}, M., {Ghez}, A.~M., {Hornstein}, S.~D., {Lu}, J.~R., \& {Morris}, M.
 2007, \apj, 659, 1241

\bibitem[{{Raghavan} {et~al.}(2010){Raghavan}, {McAlister}, {Henry}, {Latham},
 {Marcy}, {Mason}, {Gies}, {White}, \& {ten Brummelaar}}]{rag+10}
{Raghavan}, D., {McAlister}, H.~A., {Henry}, T.~J., {Latham}, D.~W., {Marcy},
 G.~W., {Mason}, B.~D., {Gies}, D.~R., {White}, R.~J., \& {ten Brummelaar},
 T.~A. 2010, \apjs, 190, 1

\bibitem[{{Rauch} \& {Tremaine}(1996)}]{rau+96}
{Rauch}, K.~P. \& {Tremaine}, S. 1996, New Astronomy, 1, 149

\bibitem[{{Remage Evans}(2011)}]{eva11}
{Remage Evans}, N. 2011, ArXiv:1102.5316

\bibitem[{{Sch{\"o}del} {et~al.}(2009){Sch{\"o}del}, {Merritt}, \&
 {Eckart}}]{S:09}
{Sch{\"o}del}, R., {Merritt}, D., \& {Eckart}, A. 2009, \aap, 502, 91

\bibitem[{{Spitzer}(1987)}]{spi87}
{Spitzer}, L. 1987, {Dynamical evolution of globular clusters} (Princeton, NJ,
 Princeton University Press, 1987, 191 p.)

\bibitem[{{Thompson}(2011)}]{tho+11}
{Thompson}, T.~A. 2011, \apj, 741, 82

\bibitem[{{Wen}(2003)}]{wen03}
{Wen}, L. 2003, ApJ, 598, 419

\bibitem[{{Willems} \& {Kolb}(2002)}]{wil+02}
{Willems}, B. \& {Kolb}, U. 2002, \mnras, 337, 1004

\bibitem[{{Willems} \& {Kolb}(2004)}]{wil+04a}
---. 2004, \aap, 419, 1057

\bibitem[{{Woosley} {et~al.}(2002){Woosley}, {Heger}, \& {Weaver}}]{woo+02}
{Woosley}, S.~E., {Heger}, A., \& {Weaver}, T.~A. 2002, Reviews of Modern
 Physics, 74, 1015

\bibitem[Wu et al.~(2007)]{WM:07}Wu, Y., Murray, N. W. \& Ramsahai, \ 2007 ApJ,  670, 820

\end{thebibliography}

\end{document}